\documentclass[11pt]{article}

\usepackage{graphicx}
\usepackage{url}
\usepackage{amsmath}
\usepackage{array}
\usepackage{textcomp}
\usepackage{MnSymbol}
\usepackage[pagebackref,pdfborder={0 0 0},colorlinks=true]{hyperref}

\newcommand{\thiswidth}{100mm}
\newcommand{\thiswidthB}{100mm}

\newcommand{\extraskip}{100mm}

\newcommand{\VideoInputDeviceCategory}{\texttt{CLSID\_Video\-Input\-Device\-Category}}
\newcommand{\cpp}{C$\plus\plus$}

\title{Video Chat with Multiple Cameras}

\author{John MacCormick\\Dickinson College Technical Report}
\date{March 2012}

\begin{document}

\maketitle

\begin{abstract}
  The dominant paradigm for video chat employs a single camera at each
  end of the conversation, but some conversations can be greatly
  enhanced by using multiple cameras at one or both ends.  This paper
  provides the first rigorous investigation of multi-camera video
  chat, concentrating especially on the ability of users to switch
  between views at either end of the conversation.  A user study of 23
  individuals analyzes the advantages and disadvantages of permitting
  a user to switch between views at a remote location.  Benchmark
  experiments employing up to four webcams simultaneously demonstrate
  that multi-camera video chat is feasible on consumer hardware.  The
  paper also presents the design of MultiCam, a software package
  permitting multi-camera video chat. Some important trade-offs in the
  design of MultiCam are discussed, and typical usage scenarios are
  analyzed.
\end{abstract}

\newpage
\tableofcontents 

\newpage
\section{Introduction}



\begin{figure}[hbtp]
  \renewcommand{\thiswidth}{70mm}
  \renewcommand{\extraskip}{5mm}
  \centering 
  \begin{tabular}{@{}%
      >{\centering}c%
      @{\hspace{3mm}}%
      >{\centering}c%
      @{}}
    \includegraphics[width=\thiswidth]{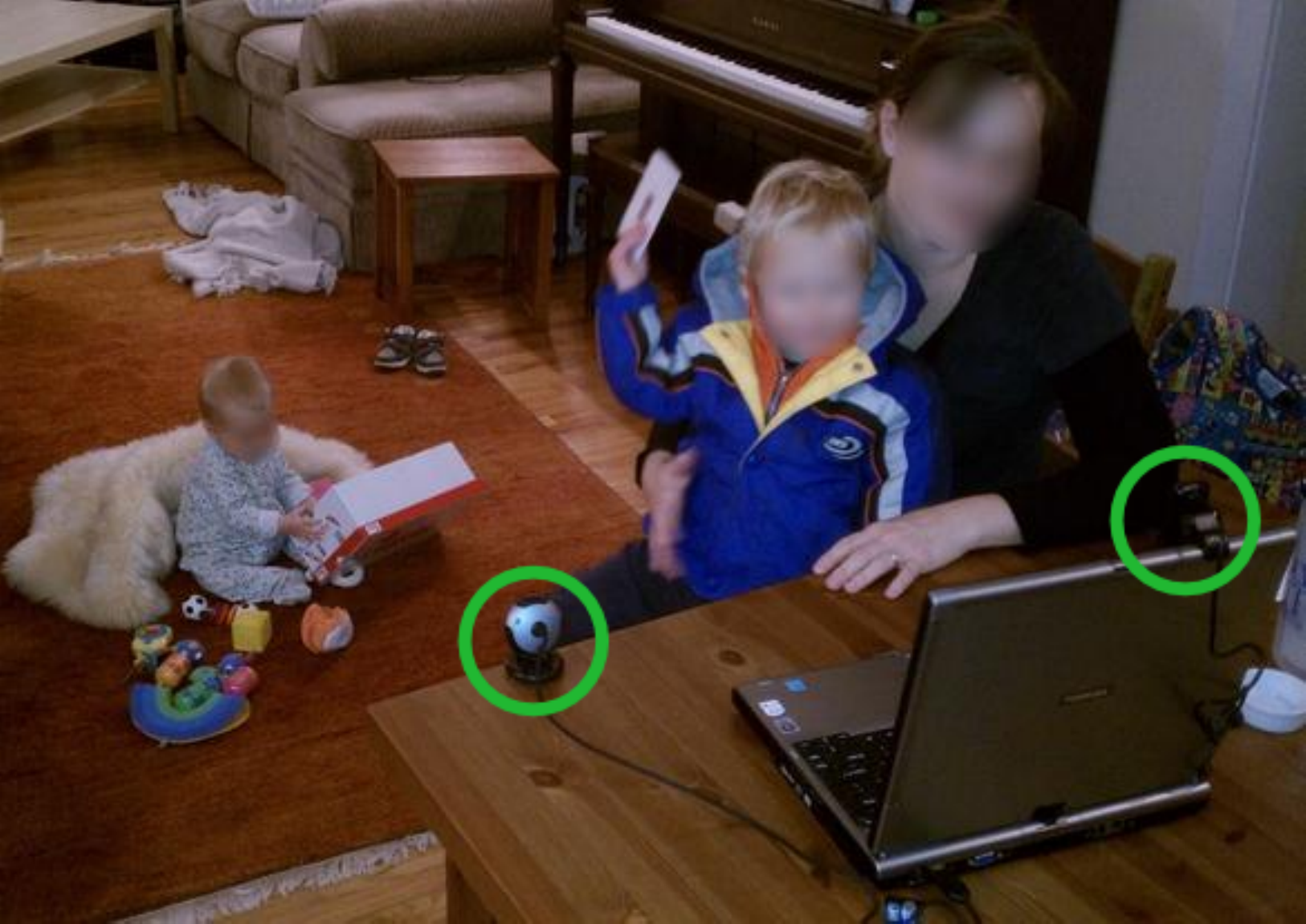}
    &
    \includegraphics[width=\thiswidth]{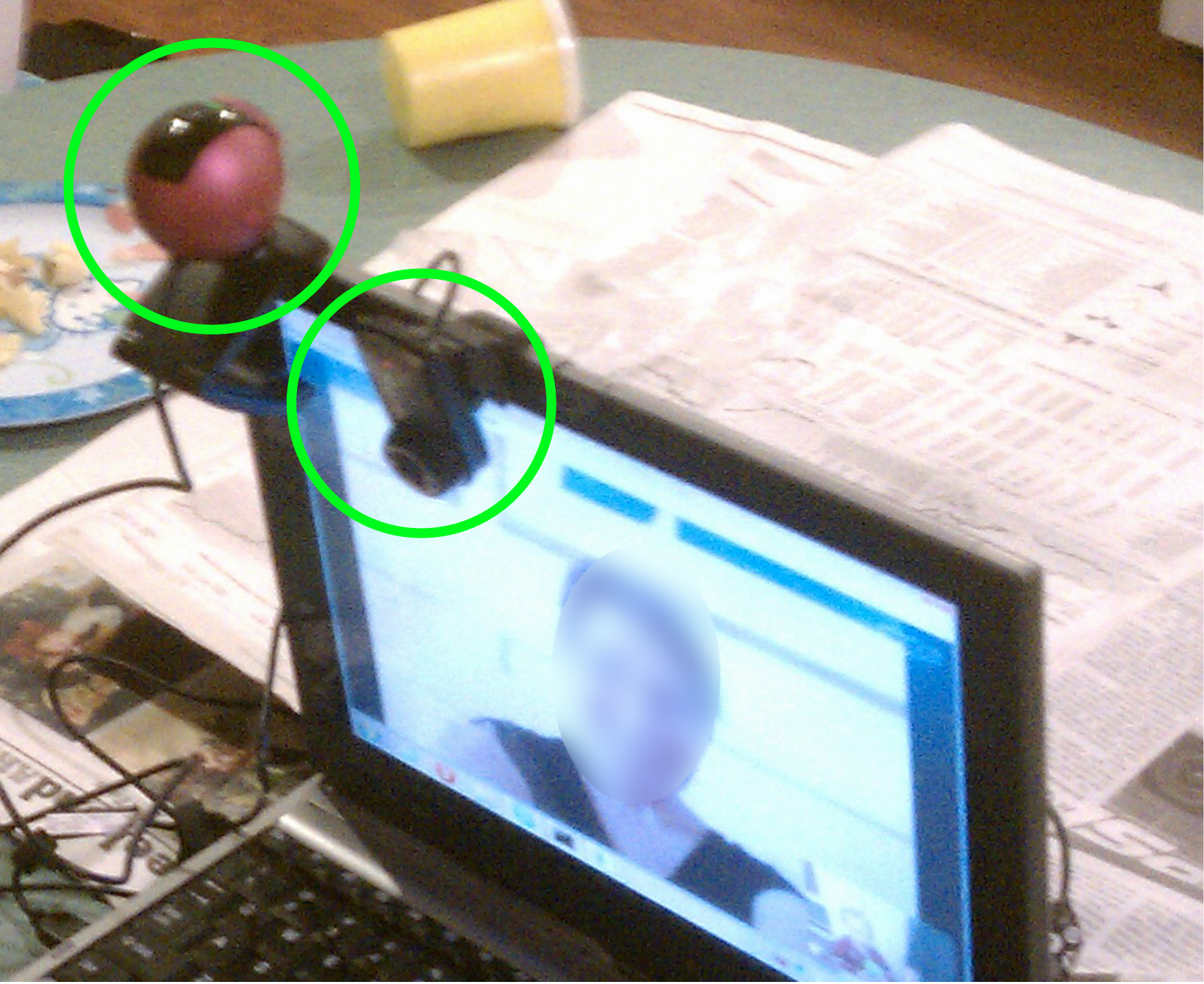}
    \tabularnewline
    (a) two forward-facing cameras
    &
    (b) forward- and rear-facing cameras
    \tabularnewline [\extraskip]
    \includegraphics[width=\thiswidth]{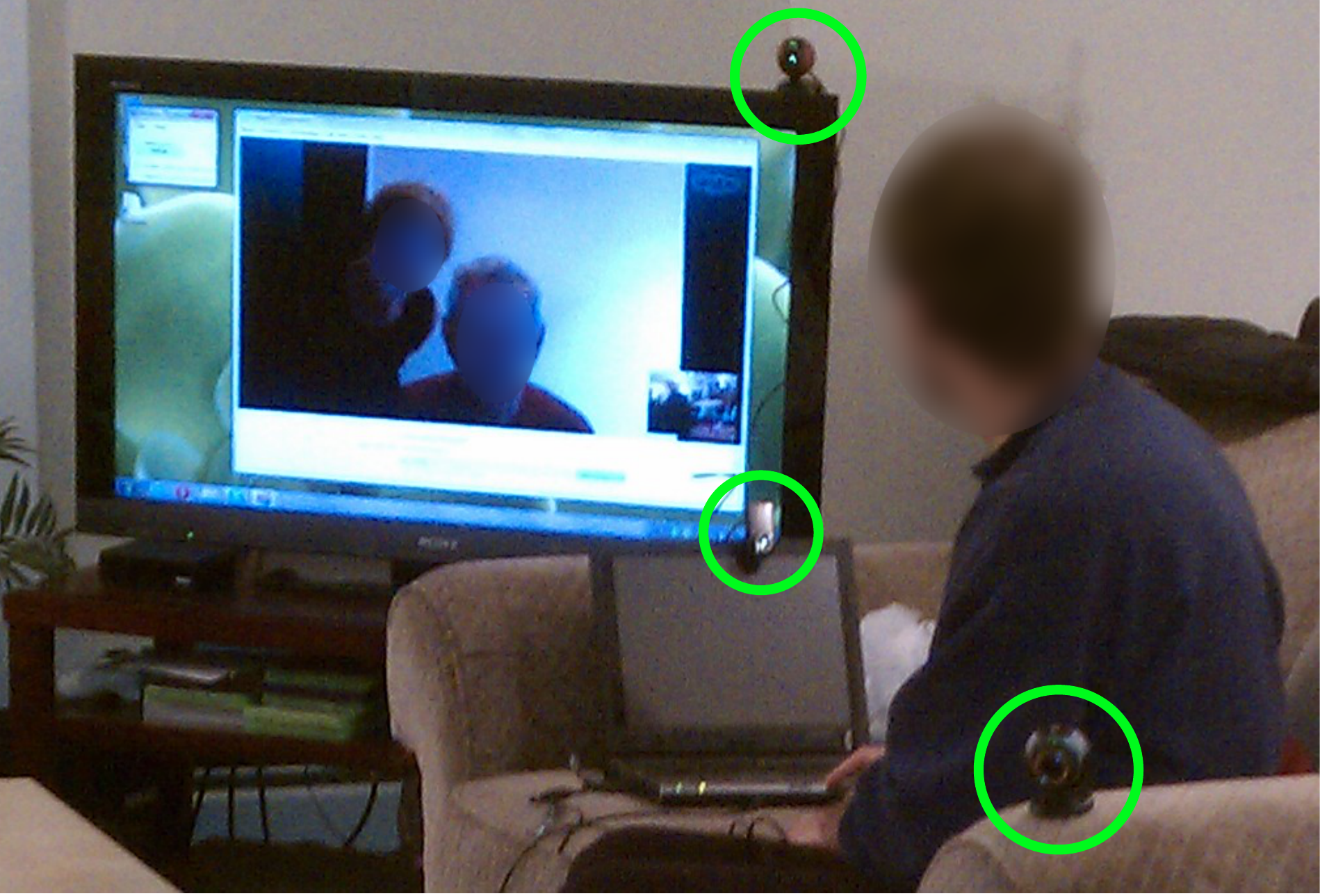}
    &
    \includegraphics[width=\thiswidth]{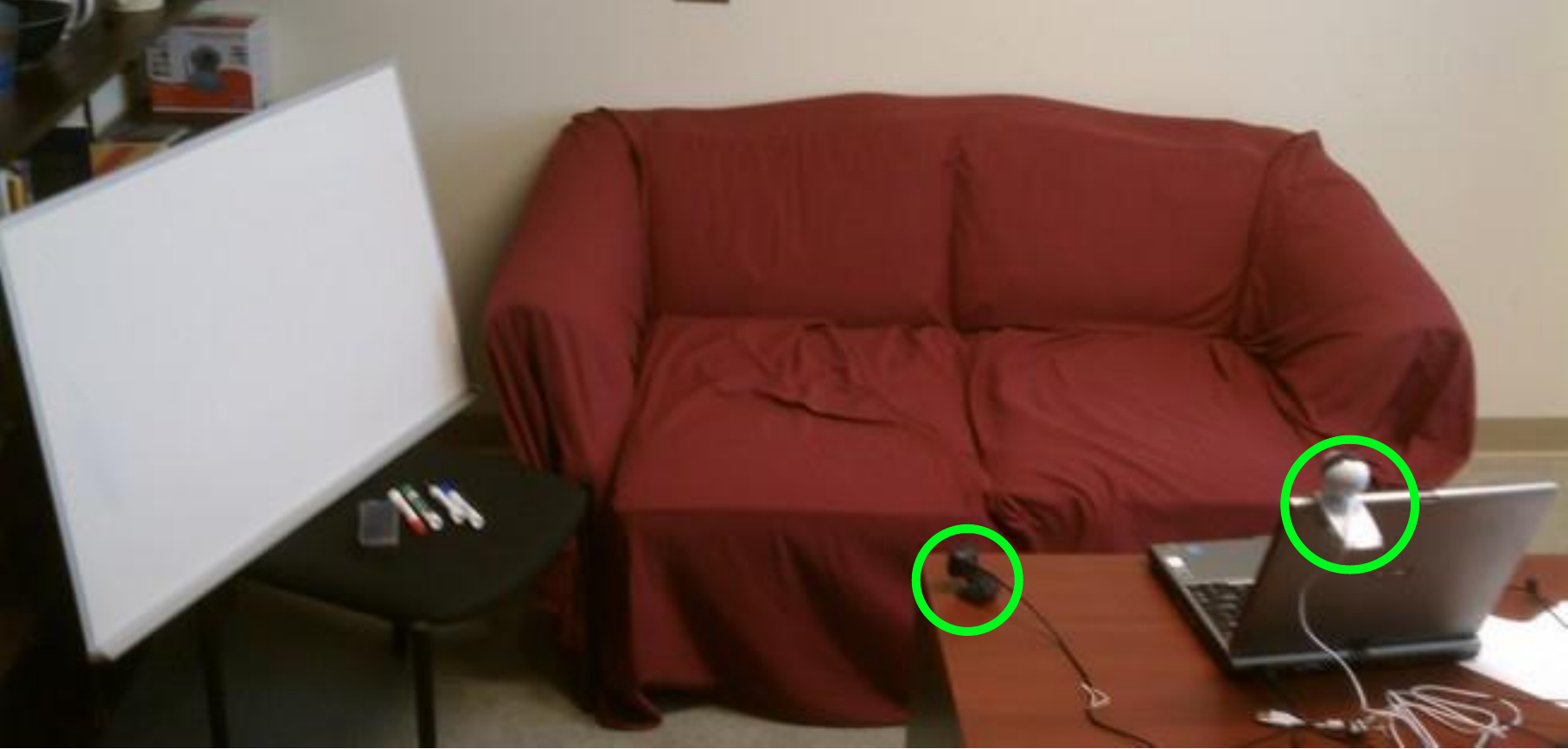} 
    \tabularnewline
    (c) wide shot, headshot, and close-up
    &
    (d) two cameras for whiteboard discussion
    \tabularnewline [\extraskip]
    \includegraphics[width=\thiswidth]{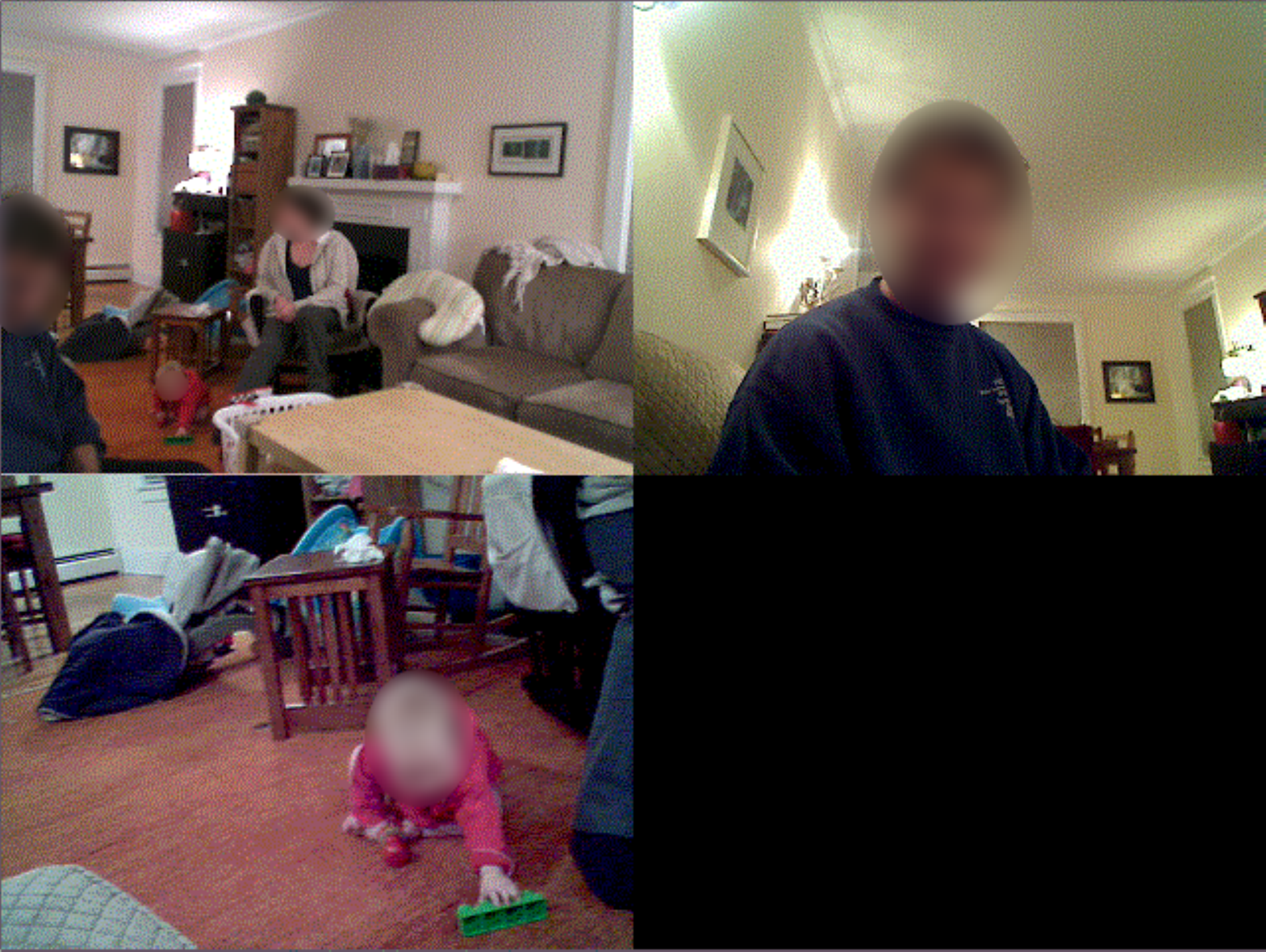}
    &
    \includegraphics[width=\thiswidth]{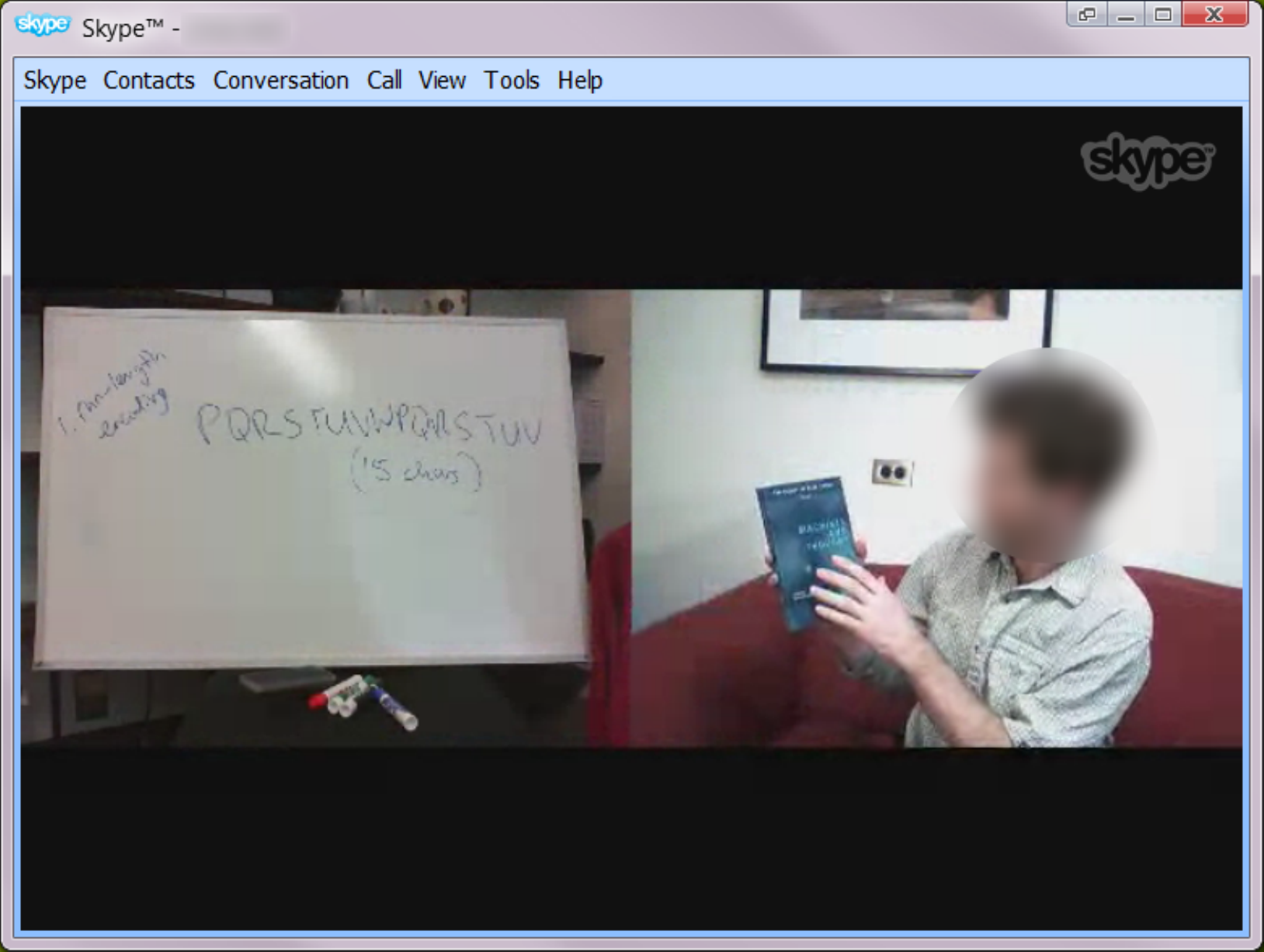} 
    \tabularnewline
    (e) remote tiled view of (c)
    &
    (f) remote tiled view of (d)
    \tabularnewline
  \end{tabular}
  \caption{\textbf{Typical MultiCam usage scenarios.} Webcams
    are highlighted by green circles.}
  \label{fig:usage-scenarios-intro}
\end{figure}

Video chat is now commonplace for a significant proportion of Internet
users, via popular, user-friendly software such as Skype, Windows Live
Messenger, Yahoo! Messenger, AOL Instant Messenger (AIM), and Google
Chat.  Skype alone reported an average of over 120~million connected
users every month in their 2010 IPO filing, and 40\% of Skype-to-Skype
chat minutes employ video~\cite{Skype2010}.  Video chat is likely to
undergo a further substantial leap in popularity with the increasing
availability of video calls on cell phones and tablets.
This report advocates and analyzes another dimension for the expansion
of video chat: the use of multiple cameras.
Figure~\ref{fig:usage-scenarios-intro} demonstrates some of the
possibilities enabled by the MultiCam software package described
later.  In each case, a laptop running Skype has two or more USB
webcams connected, and the chat participants at \emph{both} ends of
the conversation are able to switch at will between individual views
of each camera or a tiled view of all simultaneously.  The primary
goals of this report are to analyze the utility and feasibility of such
multi-camera video chats, and to discuss some important trade-offs
inherent in designing multi-camera software.

\subsection{Limitations of the single-camera paradigm}
\label{sec:limit-single-camera}

The predominant paradigm for video chat employs a single webcam at
each end of the conversation.\footnote{Evidence for this claim is
  purely anecdotal, but nevertheless seems very strong.}  For many
purposes, this is perfectly adequate.  In some cases, the
communication taking place is equivalent to a traditional audio
telephone call enhanced by the exchange of facial expressions and hand
gestures.  But the single-camera scenario also offers the opportunity
for forms of communication much further removed from a traditional
phone call.  For example, small objects can be displayed and
demonstrated by holding them up to the camera (``How do you like this
water bottle I bought yesterday?'').  An accurate impression of larger
objects can be conveyed by carrying the camera --- typically, while it
is attached to a laptop, which is also carried --- around the objects
of interest (``This is what the outdoor furniture looks like when it's
on the back patio'').  The same method of carrying camera and laptop
can be used to convey an impression of an indoor or outdoor space
(``The apartment we're staying in has this small kitchen, but as you
walk into the living room you can see this great view of the downtown
through those windows'').  Another usage pattern is to follow a moving
object such as a pet or child, again by physically moving the camera
(``Watch this --- if I walk towards her, the cat will run up the
stairs and jump into the cupboard $\ldots$ there'').  It is important
to note that in many of these usage patterns, a participant in the
chat makes use of a \textit{local view window} on the screen, which
shows the video being sent from the local camera.  This is how a
participant verifies that the remote participant can see the objects
or activities currently being discussed.  The local view window is
typically provided as a small subwindow in the main video chat window.

Despite the wide range of possible usage patterns, the single-camera
paradigm for video chat is unnecessarily restrictive and burdensome.
It is \emph{restrictive} because only a single view is available from
the single camera at any one time. It is \emph{burdensome} because the
the onus is on the person with the camera to point it at the part of
the scene that is currently of interest.

An underlying reason for these problems with the single-camera
paradigm is that the paradigm inverts one of the basic relationships
in human communication.  Some new terminology will help explain this.
At any particular instant in a conversation between two individuals,
the person who is speaking, explaining, or demonstrating an activity
or object will be referred to as the \textit{speaker}; the person
listening and watching the speaker will be referred to as the
\textit{listener}\footnote{Of course, most conversations are two-way,
  so the identities of the speaker and listener are frequently
  switched. But except for rare moments of misunderstanding, a typical
  conversation has only one speaker and one listener at any one
  instant.}.  In a face-to-face conversation, the speaker is free
to interact with the environment in whatever manner provides the most
effective communication.  Meanwhile, the listener is free to determine
which part of the scene will command his or her attention.  The
conventional video chat, on the other hand, reverses these two
freedoms.  The speaker loses the freedom to interact with the
environment, and is instead required either to move objects into the
view of the camera, or move the camera.  And in both cases, the
speaker it is often required to constantly monitor the local view
window and adjust the listener's view by moving camera or object.  The
listener also loses freedom in conventional video chat: specifically,
the freedom to choose which part of the scene is being watched. The
listener is restricted to see only what is available from the
speaker's single camera.  Of course, the freedoms described here
are not absolute.  For example, even in a face-to-face conversation,
the speaker encounters plenty of constraints, such as the social
convention to face the listener most of the time, and the necessity of
holding any objects where they can be seen. And on the other hand,
even in a standard video chat, the listener does have some freedom to
choose which part of the transmitted image will be watched most
closely.  But in both cases, there is a substantial difference between
the amounts of freedom in a face-to-face conversation and a video
chat.

As a concrete example of these restrictions, consider the following
scenario, which the author has experienced frequently. A parent and
young child are attempting a video chat with a friend or relative
using a webcam mounted on a laptop.  For concreteness, let's say the
chat is with a grandparent. The parent would like to converse with the
grandparent while also demonstrating the child's activities, which may
include crawling, climbing, running, and determined attempts to
interfere with the laptop.  This typically degenerates into 100\% of
the parent's attention being devoted to moving the laptop and camera
around, constantly checking the local view window to ensure that the
child can be seen by the grandparent.  It is very difficult to have a
satisfactory conversation under these circumstances.  Indeed, in the
author's own experience, the only viable solution for the parent is to
recruit a third person to act as full-time cameraman.  This liberates
the speaker (i.e.\ parent), but does not really free the listener
(i.e.\ grandparent) much.  For example, we can imagine that the
grandparent might wish to view the parent at certain times in this
conversation, and the child at other times.  An ideal solution would
give the grandparent freedom to switch between these views with ease.

\subsection{Scope of the report}
\label{sec:scope-report}

It may prove impossible for technology to completely restore these
face-to-face freedoms under the constraint of video chats.  But there
are three obvious avenues to explore in seeking to partially restore
these freedoms:
\begin{enumerate}
\item \label{item:multiplecameras} Employ \emph{multiple cameras}
  simultaneously, each showing a different view of the scene (thus
  reducing---but probably not eliminating---the need for the
  speaker to move cameras or objects, and monitor the local view
  window).
\item \label{item:listener-control} Permit \emph{listener control}:
  allow the listener to adjust and choose between the views offered by
  the speaker. This includes switching between cameras, viewing
  all cameras simultaneously, and could also incorporate more
  fine-grained control such as (digital or actual) pan/tilt/zoom.
\item \label{item:heterogeneous-devices} Use \emph{heterogeneous
    devices} to provide the listener with maximum choice. This could
  include standard webcams, wide-angle cameras, 3D cameras, wireless
  cameras, and panoramic cameras.
\end{enumerate}
This report investigates some aspects of
items~\ref{item:multiplecameras} and~\ref{item:listener-control}
(multiple cameras and listener control), but does not address
item~\ref{item:heterogeneous-devices}, except for a brief discussion
in Section~\ref{sec:related-work}.  And even within
items~\ref{item:multiplecameras} and~\ref{item:listener-control}, the
report examines only a small subset of the possible approaches. The
primary objective is to demonstrate the utility and the feasibility of
multi-camera video chat in general, and especially the utility of
listener control.  It is not the objective of the report to exhaustively
assess the possibilities for multiple cameras or listener control, nor
to recommend an optimum set of features.  In fact, the report is
limited to considering standard webcams only, with listener control
limited to switching between a small fixed number (2--4) of views from
these webcams.  Although this barely scratches the surface of the
potential for multi-camera chat, we will see that the report
nevertheless provides several important contributions (see
Section~\ref{sec:contribution}).

Another restriction of the scope of the report is that it specifically
addresses \emph{consumer} video chat.  Possibilities for
commercial-grade videoconferencing and professional webcasts are not
directly considered, although some of the conclusions may transfer to
those arenas.  The emphasis on consumer video chat means we seek
solutions that are widely applicable and easily used by
nonexperts. Therefore, the emphasis will be on solutions that:
(i)~involve inexpensive, standard hardware; (ii)~have moderate
computational costs (i.e.\ the software runs on a modest device
without adversely affecting a video chat); and (iii)~require only
extremely simple inputs from the user.

A final caveat is that the report does not seek to quantify the
benefits of multi-camera video chat, when compared to the
single-camera approach. As already stated above, it seems clear that
the single-camera paradigm is perfectly adequate for many video chats.
On the other hand, it is equally obvious that some scenarios can
benefit from multiple cameras.  Any attempt to quantify these benefits
suffers from a severe chicken-and-egg problem.  To see this, suppose
that a hypothetical experiment reached the following conclusion: ``5\%
of single-camera video chats would be significantly enhanced by using
multiple cameras, where `significantly enhanced' means an increase in
measured user satisfaction of at least 30\%.''  This hypothetical
conclusion might seem disappointing, since only 5\% of chats benefit
significantly.  But of course, the experiment would be conducted
within the present ecosystem of single-camera chats. So the population
of chats sampled by the experiment would be heavily biased towards
setups constructed with the single-camera paradigm in mind.  It is
possible that if multi-camera tools were widely available and easily
used, the video chat ecosystem would be altered significantly and a
much larger proportion of video chats could benefit.  Hence, the
approach of the present report is to demonstrate that compelling
multi-camera scenarios exist (see next paragraph), without addressing
the question of whether such scenarios comprise a substantial fraction
of video chats---according to the chicken-and-egg argument above, this
question is irrelevant.

Let us immediately discuss the existence of compelling multi-camera
scenarios for consumer video chat. There are four types of evidence
here.  First, the report explicitly describes two scenarios that are
difficult or impossible without multiple cameras: the ``children in
the background'' scenario of Section~\ref{sec:typical-usage}, and the
``whiteboard lecture'' scenario of Section~\ref{sec:user-study}.
Second, there are several existing systems that offer multi-camera
consumer video chat (see Section~\ref{sec:related-work} for details).
I am not aware of any published statistics on the usage of such
systems, but the mere existence of these systems does suggest there is
demand for them.  Third, I do have download statistics for the
MultiCam software package introduced in this report: despite being a
relatively immature research prototype with no publicity beyond a
posting to some Skype forums, MultiCam is being downloaded dozens of
times per month at the time of writing.  Fourth, some commercial
videoconferencing systems offer certain multi-camera capabilities (see
Section~\ref{sec:related-work} for examples), although not in the
chat-friendly format suggested here, as far as I am aware.
Presumably, consumers would also like these capabilities if they were
available at little or no cost.


\section{Overview of MultiCam usage}
\label{sec:MultiCam-overview}

The experiments described later employ a software package, called
MultiCam, written by the author specifically for this research. The
MultiCam software itself is not a primary contribution of the report,
although it does have some limited novelty, as described in
Section~\ref{sec:related-work}.  Nevertheless, it will be useful to
understand MultiCam's functionality before discussing related work
(Section~\ref{sec:related-work}) and the new results of this report
(Sections~\ref{sec:user-study} and~\ref{sec:benchmarks}).  Therefore,
this section gives an overview of MultiCam from the point of view of a
\emph{user}.  The design of MultiCam, from the point of view of a
\emph{programmer}, is discussed separately, in
Section~\ref{sec:MultiCam-design}.  

MultiCam is free and open source, and is available for Microsoft
Windows only (more precisely, Windows~7 and later).  The local
camera-switching functionality of MultiCam works, in principle, with
any video chat software.\footnote{In practice, MultiCam has certain
  technical problems with some video chat software.  The present
  implementation works well on Skype, Yahoo!\ Messenger and ooVoo, for
  example, but not on Google Chat.} Remote camera-switching, on the
other hand, works only with Skype, since it relies on Skype's
so-called \emph{desktop API}.  For concreteness, the remainder of the
report focuses on running MultiCam with Skype only.

\begin{figure}
  \centering
  \includegraphics[width=80mm]{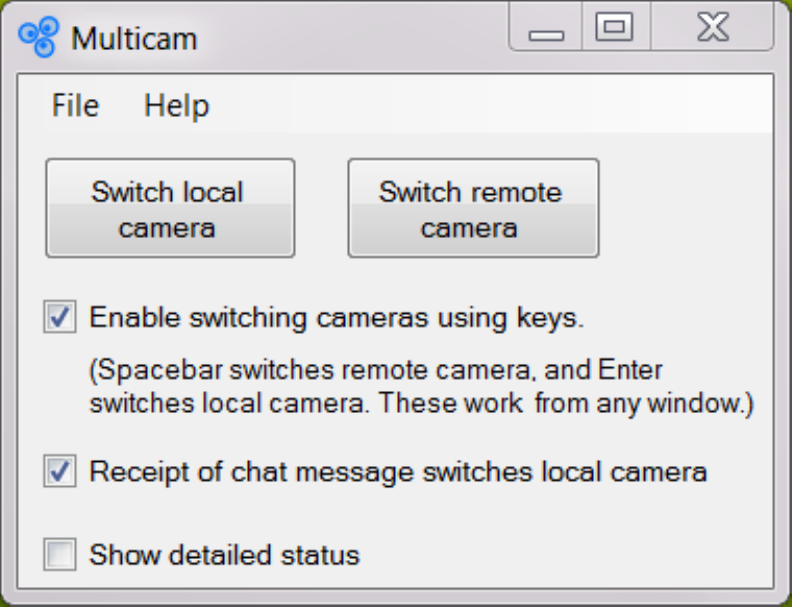}
  \caption{\textbf{Screenshot of the MultiCam application.}}
  \label{fig:MultiCam-app-screenshot}
\end{figure}
\begin{figure}
  \centering
  \fbox{\includegraphics[width=388pt]{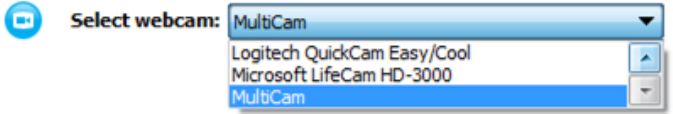}}
  \caption{\textbf{Screenshot demonstrating selection of the MultiCam
      virtual camera as the video input for Skype.}}
  \label{fig:MultiCam-select}
\end{figure}

MultiCam consists of two components: the MultiCam application, and the
MultiCam virtual camera.  The MultiCam application, shown in
Figure~\ref{fig:MultiCam-app-screenshot}, is a stand-alone GUI
application that allows the user to adjust settings and to perform
camera-switching functions during a video chat.  The MultiCam virtual
camera appears, to the operating system, to be a standard video camera
device. Video chat software such as Skype therefore offers
``MultiCam'' as one of the options when a user selects a video input
device, as shown in Figure~\ref{fig:MultiCam-select}.  In reality, of
course, the MultiCam virtual camera is not a physical camera. Instead,
it multiplexes the machine's physical cameras: it passes video data
from one or more of the physical cameras to the video chat software,
possibly after transforming the data in some way.

\begin{figure}
  \renewcommand{\thiswidth}{58mm}
  \centering
  \begin{tabular}{@{}c@{\hspace{4mm}}c@{}}
    \includegraphics[width=\thiswidth]{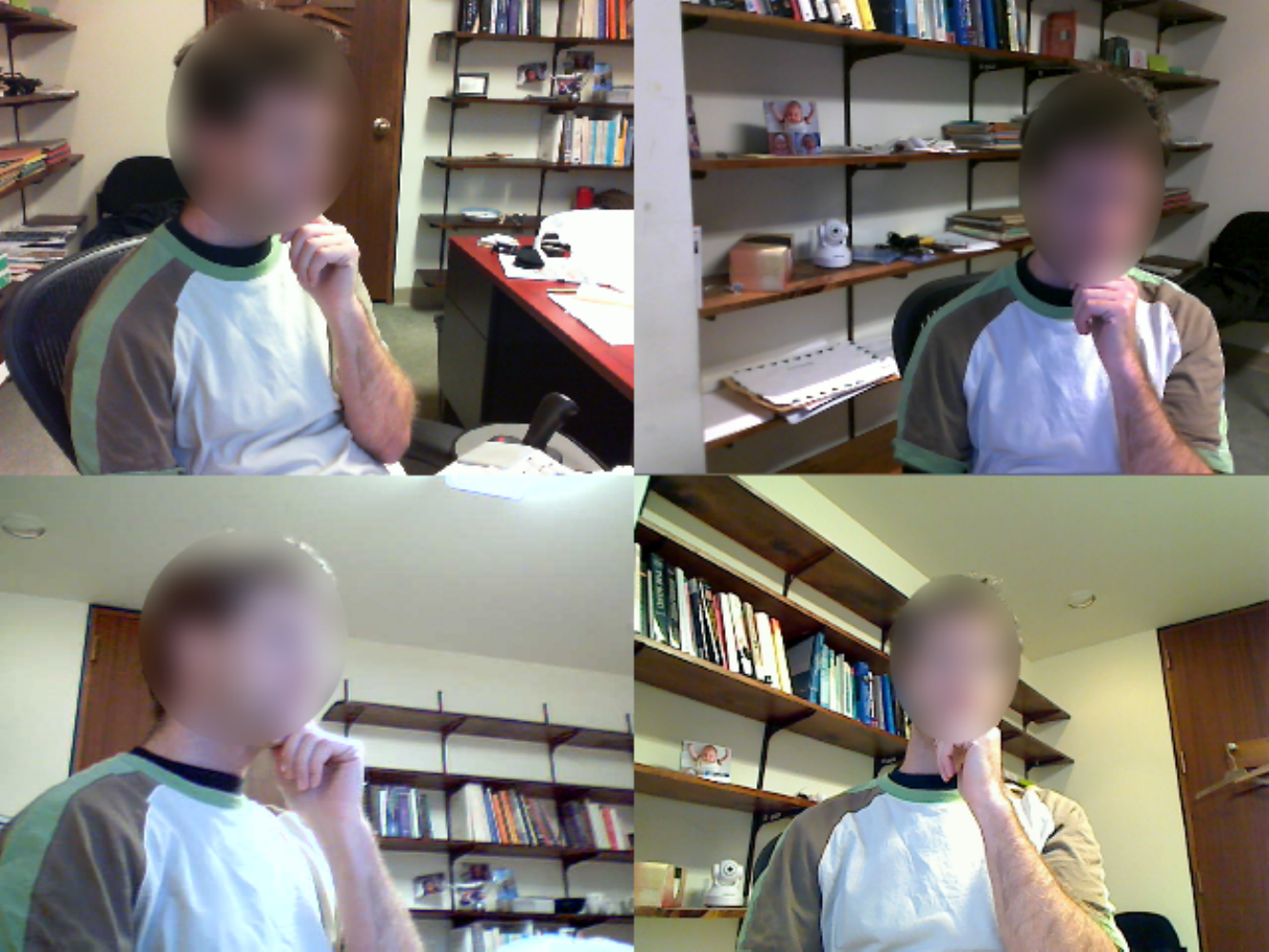}
    &
    \includegraphics[width=\thiswidth]{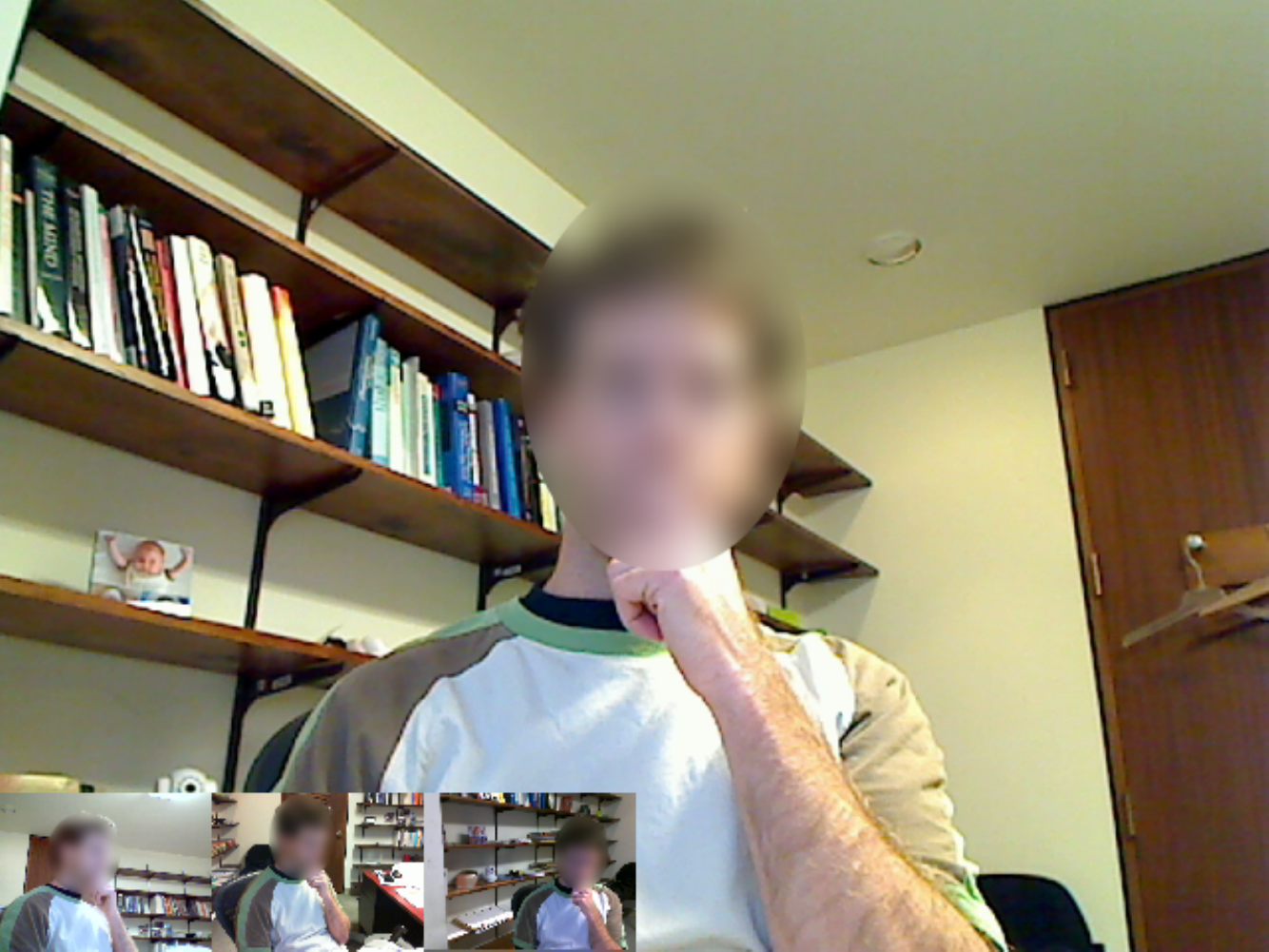}
    \\
  \end{tabular}
  \caption{\textbf{MultiCam's two modes.}  Left: In the tiled mode,
    the virtual camera shot contains subsampled versions of frames
    from each physical camera. Right: In the non-tiled mode, the
    virtual camera fills the frame with the output from one physical
    camera, then adds small overlays from the other cameras.}
  \label{fig:tiled}
\end{figure}

To be more specific, MultiCam has two high-level modes: \emph{tiled},
and \emph{non-tiled}---these are shown in Figure~\ref{fig:tiled}.  The
tiled mode places subsampled versions of the input from each physical
camera into a single output image.  When in non-tiled mode, one of the
physical cameras is designated by the user as the \emph{primary
  camera}.  The input from the primary camera is transferred unaltered
to the output image, but some small subsampled versions of the other
(non-primary) physical cameras are overlaid at the bottom left of this
output.  The identity of the primary camera is not fixed.  Indeed, the
MultiCam application permits users to switch the identity of the
primary, and to switch between tiled and non-tiled modes, with a
single keystroke or mouse click (see
Figure~\ref{fig:MultiCam-app-screenshot}).

This brings us to the most important design decision for the MultiCam
UI: how should the user switch cameras?  Clearly, it is desirable that
the user can switch cameras quickly and easily, preferably with a
single keystroke or mouse click. On the other hand, some early
experimentation demonstrated that control of multi-camera chats can be
bewildering. Recall that there may several (say, up to four) cameras
at each end of the conversation, and each user can control the
camera-switching at both ends.  So a user may be faced with as many as
eight possibilities when switching views. To reduce the possible
confusion to an absolute minimum, MultiCam adopts an extremely limited
interface.  There are only two actions a user can perform:
\emph{advance} the local camera, or \emph{advance} the remote camera.

The word ``advance'' here has a specific technical meaning, defined as
follows.  The $N$ cameras connected to any one machine have a natural
ordering, determined by the order in which the operating system
enumerates them. So they can be the assigned numerical IDs $1,2,\ldots
N$.  If the system is in non-tiled mode when the user advances the
camera, the ID of the primary camera is incremented by one---so we
literally ``advance'' to the next camera's view.  Of course, there is
an exception if the primary camera ID is already equal to $N$: in this
case, the system switches into tiled mode. Finally, if the system is
currently in tiled mode, it switches to non-tiled with primary camera
ID equal to~1.  

Thus, the user cycles through the $N+1$ possible views in a fixed
order. There is no way to jump directly to a particular desired view.
The advantage of this is that control by keystroke remains
feasible---and this is especially important when we consider that the
user may want to maximize the video chat window, leaving the MultiCam
application window invisible. In the current implementation, for
example, the user hits the Enter key to advance the local camera, and
the spacebar to advance the remote camera.  Even for unpracticed
users, memorizing these two keystrokes is easily achievable.  On the
other hand, there is some evidence from the user study
(Section~\ref{sec:user-study}) that users would prefer to have a
method of jumping directly to the desired view. Investigation of this
is left for future work, primarily because of the difficulty of
integrating any sensible solution with existing video chat
software. For example, one plausible possibility would be for a user
to switch views by clicking on the thumbnail of the desired new
primary camera (or a suitable overlaid icon for tiled mode).  But this
would require MultiCam to capture mouse clicks from the video chat
window, which is not directly supported by existing video chat
software.  In any case, this solution is still unsatisfactory for
chats with multiple cameras at both ends, since the number of
thumbnails on the screen may be overwhelming for the user. There are
some very interesting possibilities for future work here, discussed in
Section~\ref{sec:discussion}.

Naturally, the keystroke-switching feature can be disabled, so the
user can employ the keyboard for some other purpose while chatting, if
desired.  When not using keystrokes, a user can still switch cameras
by clicking on the ``Switch local camera'' or ``Switch remote camera''
buttons in the MultiCam application window.

MultiCam permits one additional method of switching cameras.  When the
MultiCam application is running, any Skype instant message received
advances the local camera setting.  Clearly, this is an abuse of the
intended functionality for Skype instant messages, but there is a good
reason for incorporating this instant message (IM) hack.  To
understand this reason, suppose user $A$ is video Skyping with user
$B$.  Suppose further that $A$ has multiple cameras and is running the
MultiCam application, whereas $B$ is not running MultiCam. Without the
IM hack, $B$ would have no method of advancing $A$'s camera in this
situation. 

The key point is that IM is an integral part of all Skype clients and
is guaranteed to be available to any user, whereas an application such
as MultiCam (or even a more tightly-bound plugin) requires a separate
installation procedure before a user can enjoy the benefits of remote
camera switching.  The ability to switch cameras via IM is especially
crucial for Mac and Linux users, since the MultiCam application is
Windows-only at the time of writing.  
As with switch-by-keystroke, the switch-by-IM feature can be disabled,
which is obviously necessary if the chat participants wish to use IM
for its intended purpose.

Finally, note that MultiCam offers a small thumbnail of all
non-primary views when in non-tiled mode.  An example of this can be
seen in the right-hand panel of Figure~\ref{fig:tiled}: in the bottom
left of the panel, we see thumbnails of three other views.  Early
experience with MultiCam suggested these thumbnails are helpful, but
they also add clutter.  These thumbnails might even overload the user
with information. This is especially true given that typical video
chat software also displays yet another thumbnail, showing the remote
participant's view of the local scene.  To make this concrete,
consider a specific example, in which users $A$ and $B$ are chatting
via Skype and MultiCam.  Suppose user $A$ has two cameras, denoted
$A_1, A_2$, and user $B$ has three cameras, denoted $B_1, B_2, B_3$.
Suppose further that $A$ has selected non-tiled mode with $B_1$ as
primary, and $B$ has selected non-tiled mode with $A_1$ as primary.
Then with default Skype and MultiCam settings, $A$'s view consists of
a large view of $B_1$, overlaid with small thumbnails of $B_2$ and
$B_3$ in the bottom left, and a medium-sized thumbnail (showing $B$'s
view) of $A_1$ in the bottom right, but this latter thumbnail has an
additional sub-thumbnail of $A_2$ overlaid in its own bottom left.
Clearly, this could be a case of too much information for the user,
and future work should investigate cleaner ways of conveying it.

\section{Related work}
\label{sec:related-work}

In this section, we survey two strands of related work: (i)
multi-camera video chat, and (ii) more immersive telepresence
projects.  It is claimed that this report occupies a vacant niche in
the literature, because academic projects and publications have
focused on (ii), whereas this report focuses on the simpler ideas in
(i).  More specifically, software for (i) has been available for at
least a decade,\footnote{In a 2011 Skype forum posting about switching
  between multiple cameras (\url{http://tinyurl.com/skypeforum}), a
  user stated ``i've been using manycam (and a predecessor) for a
  decade!''} but the utility and feasibility of such
software---especially the possibility of listener-controlled
camera-switching---has not been rigorously analyzed. This report
provides that analysis.

\subsection{Related multi-camera video chat software and hardware}
\label{sec:related-multi-camera}

Several existing software products offer convenient ways for the
speaker to switch between cameras during video chat. These
include ManyCam\footnote{\url{http://www.manycam.com}} and
WebcamMax.\footnote{\url{http://www.webcammax.com}} In addition to
easy switching, both these products permit a picture-in-picture
mode, in which one camera's output can be overlaid on the
other's---this is a similar to MultiCam's tiled mode.  Neither
offers simultaneous viewing of more than two cameras, nor do they
permit listener-controlled switching.  To the best of my knowledge,
the design of these tools is not discussed in any
publicly-disclosed documents.  But both are marketed primarily as
webcam effects software: that is, as a tool for altering the
output of a single, primary camera.  ManyCam is also prominently
marketed for simultaneously using a single webcam in multiple
applications, in contrast to MultiCam's primary goal of using
multiple webcams in a single application.  The differing emphases
of MultiCam and these existing software products will be important
during the discussion of MultiCam's design in
Section~\ref{sec:design-overview}.

Another existing alternative is VH MultiCam
Studio\footnote{Discontinued, but still available from
  \url{http://www.mediafire.com/?nxzvrhzzzkz} at the time of
  writing.}  (VHMS).  VHMS permits tiling of an arbitrary number
of cameras, and even allows listener-controlled switching in Skype
via a variant of the IM hack described in
Section~\ref{sec:MultiCam-overview}, but the interface relies on
saving a number of preset configurations and would be challenging
for novice users.  The three products just mentioned are
closed-source.

It is worth noting that even with the present version of Skype~(5.5),
it is possible to use multiple cameras without resorting to virtual
camera software such as ManyCam. One approach is to launch separate
instances of Skype on the same machine,\footnote{This can be done with
  the \texttt{secondary} commandline argument.} log in with different
identifiers on these instances, assign different cameras to the
instances, and start a multiparty chat.  Obviously, this
involves some inconvenience, and consumes more bandwidth than
necessary.  Alternatively, one can use a single instance of Skype and
use the existing UI to switch between cameras during a chat. This
method requires several clicks, and generally involves several seconds
of latency after the new camera has been selected, but is by no means
unusable.

A relatively recent development is the emergence of mobile devices and
tablets with two cameras (e.g.\ Apple's iPad~2, HTC's Droid
Incredible~2).  These devices have one camera on the front, intended
primarily for video chat; and one on the back, intended primarily for
capturing photos and video.  But of course it is possible to use both
cameras during video chat, and some chat clients already support this
at the time of writing (e.g.\ Google Talk, Skype Mobile).  These
clients support convenient, intuitive speaker-controlled
switching between the two cameras.  However, they do not support
simultaneous views of both cameras, nor do they support
listener-controlled switching.


Although this report focuses on two-way video chat, some of the
multi-camera discussion also applies to the scenario of
\emph{webcasting}---a type of one-way communication in which the
speaker broadcasts to multiple listeners, who may or may not be
watching in real time.  Webcasting involves two separate tasks:
content creation (typically done using specialized webcasting software
such as CamTwist Studio, Webcam Studio, or WebCaster), and content
delivery (typically via online platforms such as Ustream or
Livestream).  Only the former concerns us here.  Webcasting software
focuses on fusing multiple types of media such as screenshots,
webcams, presentation slides, and prerecorded video clips---but the
software often also permits multiple webcams as inputs. And because
webcasting software is typically implemented via the same virtual
camera technique used in multi-camera video chat software, one can
therefore use it for a multi-camera video chat.  This may even be the
most effective choice for a well-practiced user who wishes to use the
sophisticated features of webcasting software.  But this is less than
ideal for the simple camera-switching envisaged in the current
report. Moreover, to the best of my knowledge, no webcasting software
permits listener-controlled switching between cameras.

An obvious alternative to switching between multiple cameras is to
arrange for a single camera to move via pan/tilt/zoom (PTZ).  There
are many interesting possibilities here, including the use of motion
detection~(e.g.~\cite{huang02:pan-tilt}) or a physical sensor located
on the target of interest
(e.g. Swivl\footnote{\url{http://www.swivl.com}}).  These approaches
are complementary to the focus of this report.  Of more direct
relevance are the possibilities for remote (i.e.\ listener) control of
PTZ cameras.  At the time of writing, IP cameras with remote control
functionality are available for under \$100, which comes close to our
goal of being usable for consumer video chat. Unfortunately, present
IP cameras are typically designed for surveillance applications; their
image quality tends to be inadequate for enjoyable video chat, and the
remote control interfaces can be clunky.  In any case, such cameras
can still be regarded as complementary to this report's proposals.
Even if low-cost, high-quality, remote-controlled PTZ cameras with
slick interfaces were available, we can still imagine enhancing the
video chat further by using several such cameras simultaneously with a
MultiCam-like camera-switching interface layered on top.

Although outside the scope of this report, it's important to realize
that multi-camera video chat could be enhanced by non-standard
cameras.  One simple but liberating possibility is the use of wireless
cameras.  Surprisingly, at the time of writing (March 2012), there is
no Bluetooth camera suitable for consumer video chat available for
Windows systems, and only one such camera for Apple systems (Ecamm's
BT-1).  Wireless IP cameras are another option, in principle, but
appear to be designed primarily for surveillance purposes and the
consumer-grade versions typically have poor image quality.  Smartphone
cameras can be converted into wireless webcams via apps such as
DroidCam\footnote{\url{http://www.dev47apps.com/droidcam}} and
SmartCam.\footnote{\url{http://smartcam.sourceforge.net}}  This is a
very promising approach; the only disadvantage is that the user must
either own an adjustable smartphone tripod, or manually hold the
smartphone in position.  Presumably, the ecosystem of
consumer-friendly wireless webcams will expand significantly in the
near future.

Panoramic cameras represent another alternative for enhancing video
chat. These have been previously explored in academic research
projects such as FlyCam~\cite{Foote-FlyCam-2000}, and are now
available as relatively inexpensive consumer products such as the
GoPano micro.\footnote{\url{http://www.gopano.com}} It would be very
interesting to combine this product with listener control and
multi-camera switching.

In contrast to all the above alternatives, the MultiCam software
presented in this report offers single-keystroke (or mouse-click)
switching by both speaker and listener, between an arbitrary
number of cameras, and including a tiled mode. Hence, there is a
small amount of novelty in the software itself, especially given
that MultiCam is open source.  But the minutiae of features
offered by such software are fairly unimportant, given that these
features are all easy to add.  Video chat software makers such as
Skype, for instance, could easily incorporate all of MultiCam's
functionality directly into their products with an insignificant
investment of perhaps one or two programmer-months.  Indeed, I
hope that after reading this report, video chat software makers
will include these features, taking account of the findings
described in later sections.

\subsection{Related immersive telepresence projects}
\label{sec:related-immersive}

The goal of this report is related to, but separate from, the goal of
immersive telepresence.  In this report, we seek to enhance the
listener's experience by providing multiple views of the speaker's
location, and by giving the listener control over switching between
those views. In contrast, immersive telepresence seeks to enhance the
listener's experience by creating the impression that the listener is
immersed in the speaker's location (or perhaps a virtual location
instead).  For example, the goal of the BiReality
system~\cite{Jouppi:2004:BiReality} is ``to recreate to the greatest
extent practical, $\ldots$ the sensory experience relevant for
face-to-face interactions,'' by immersing a physical robot in a remote
location.  The 3DPresence project~\cite{divorra10:_towar_aware_telep}
recreates some of the important 3D and perspective effects in a
physical conference-room setting, surpassing the immersiveness of
existing commercial telepresence systems such as Cisco Telepresence
and HP Halo.  Numerous other projects, such as
ViewCast~\cite{Yang:2007:acmmm:viewcast} and Coliseum~\cite{Baker2003},
place all participants into a virtual environment.  A survey by Otto
\textit{et al.}~\cite{Otto:2006:immersiveCVE} gives further examples
focused on collaboration by geographically dispersed users.

Implicit in all these projects is the assumption that the quality of
the listener's experience will increase with the extent and fidelity of
the immersiveness.  This assumption may be true in general---and is
particularly apt for certain facets of communication such as gaze and
gesture awareness~\cite{Kleinke86}---but it does not preclude
improving the listener's experience through other, simpler means. The
goal of this report is to do just that: without seeking immersiveness,
we can give the listener more options and more control via the much
simpler strategy of employing multiple views.



\subsection{Contribution}
\label{sec:contribution}

The two primary contributions of the report are: 
\begin{itemize}
\item It demonstrates the \emph{utility and feasibility of multi-camera
  video chat} for certain applications.
\item It analyzes the desirability of \emph{remote control over the
  camera view}.
\end{itemize}
To the best of the author's knowledge, no previous publication has
addressed these points in detail.
The secondary contributions of the report are:
\begin{itemize}
\item It describes the design trade-offs inherent in building virtual
  camera software to multiplex several cameras simultaneously, and
  offers an open-source solution to this problem.
\item It makes recommendations for a standardized protocol that could
  be used by future multi-camera software modules to interact with
  video chat software.
\item It identifies several areas in which webcam manufacturers and
  video chat software developers could enhance their support of
  multi-camera use.
\end{itemize}
Again, to the best of the author's knowledge, the existing literature
does not address any of these contributions in detail.


\section{Design of MultiCam}
\label{sec:MultiCam-design}

Section~\ref{sec:MultiCam-overview} described the design of MultiCam
from the point of view of a user.  In this section, we look under the
hood, discussing design decisions taken by the programmer. We will
concentrate on design issues with direct relevance to the user study
(Section~\ref{sec:user-study}) and benchmark experiments
(Section~\ref{sec:benchmarks}), relegating other details to
Appendix~\ref{sec:mult-design-deta}.

\subsection{Design overview}
\label{sec:design-overview}

\begin{figure}
  \centering
  \includegraphics[width=120mm]{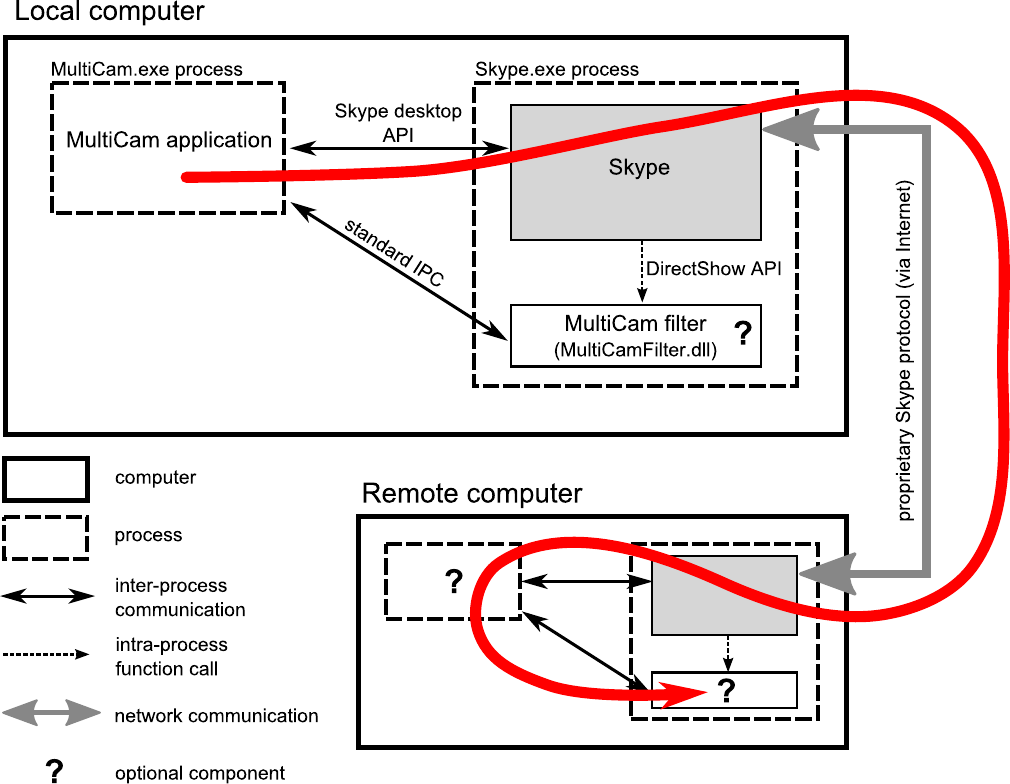}
  \caption{\textbf{Communication between MultiCam components.} The red
    arrow shows the route taken by an ``Advance remote camera''
    request issued by a local MultiCam instance.  The MultiCam
    application communicates with the Skype application via the
    publicly-documented Skype desktop API, and with the MultiCam
    filter via standard IPC. Skype issues standard DirectShow camera
    commands to the MultiCam filter by making function calls, just as
    with a physical camera.  Two instances of Skype communicate over
    the Internet via Skype's proprietary protocol.  Two instances of
    the MultiCam application also communicate with each other via the
    Skype desktop API, which transparently employs the proprietary
    link between Skype instances.  The question marks indicate
    components that may be absent.}
  \label{fig:communication}
\end{figure}

As already discussed, the MultiCam software consists of two largely
independent modules: the MultiCam \emph{application}
(\texttt{MultiCam.exe}) and the MultiCam \emph{virtual camera}
(\texttt{MultiCamFilter.dll}).  For reasons that will become clear
shortly, the MultiCam virtual camera is more accurately termed the
MultiCam \emph{filter}, and we use the latter term for the remainder
of this section.  Figure~\ref{fig:communication} gives an overview of
how the two components communicate with each other and with Skype.  As
we see in the Figure, four different types of communication are used:
standard Windows interprocess communication (IPC), API calls from the
Windows DirectShow framework~\cite{pesce03:directshow}, the
publicly-available \emph{Skype desktop API}~\cite{SkypeAPI2011}, and
the proprietary Skype protocol.  Each of these is described in more
detail below. As a concrete example, the red arrow in
Figure~\ref{fig:communication} shows the chain of communication that
occurs when a user clicks on ``Advance remote camera'' in the MultiCam
application; this too is discussed in more detail below.

We previously saw, in Section~\ref{sec:MultiCam-overview}, that the
MultiCam application is a stand-alone GUI application that allows the
user to adjust settings and to perform camera-switching functions
during a video chat (see Figure~\ref{fig:MultiCam-app-screenshot}).
But the MultiCam application has a second crucial role: it monitors
the state of other components in the system and takes actions based on
its observations.  More specifically, the components monitored by the
MultiCam application are: the local instance of Skype, the local
MultiCam filter, and the remote MultiCam filter.  Each of these
components may be present or absent, active or inactive, and has
certain additional state.

The MultiCam filter is a dynamic link library (DLL), and we have
already seen that it acts as a virtual camera, multiplexing the
machine's physical cameras.  The MultiCam filter is implemented in the
Windows DirectShow framework~\cite{pesce03:directshow}, which is a
Microsoft framework for multimedia programming.\footnote{DirectShow is
  being phased out in favor of Microsoft Media
  Foundation~\cite{polinger11}.  At the time of writing, however,
  Media Foundation is not sufficiently well-supported to form the
  basis for MultiCam.} In this framework, modules that create,
consume, or transform multimedia data are known as \emph{filters};
hence the choice of the terminology ``MultiCam filter.''

The chief abstraction in DirectShow is a directed graph known as a
\emph{filter graph}. The edges on the graph represent paths along
which video and audio data can flow, and vertices of the graph
represent filters.  In DirectShow, filters are implemented as \cpp\
classes derived from a base filter class.  The standard DirectShow
library provides numerous more specialized filter types; of particular
interest to us are the \emph{source filter} and the \emph{transform
  filter}.  A source filter has no inputs: it creates and outputs its
own multimedia data.  A webcam (or, more accurately, its device
driver) is one example of a source filter.  In contrast, a transform
filter ingests multimedia data from one or more inputs, combines or
alters those inputs in some way, then outputs the result.  Common
transform filters include those that convert between color spaces or
alter the aspect ratio of a video stream.  Of course, implementors can
write their own filters, and they are free to blur the boundaries
between filters in the standard library.

This is exactly what the MultiCam filter does. Specifically, the
MultiCam filter is implemented so that it appears, from the point of
view of any video chat software, to be a source filter.  The video
chat software can insert the MultiCam filter into a DirectShow filter
graph, and begin extracting video frames from the filter when it is
ready to do so.  In reality, the MultiCam filter is not a source
filter but a transform filter.  When the MultiCam filter detects that
it has been added to a DirectShow filter graph, it immediately creates
some new vertices in the graph---one for each physical camera in the
system---and creates connections from the new vertices to itself.  

Thus, the MultiCam filter has access to the latest image frame from
each physical camera.  It is a simple matter to reassemble these into
the desired output.  Recall from Section~\ref{sec:MultiCam-overview}
that the current filter state is specified by the ID of the primary
camera, and value of the tiled/non-tiled flag.  Thus, the output is
assembled by subsampling and shifting the inputs if necessary, then
placing them in the filter's output buffer, producing frames such as
those shown in Figure~\ref{fig:tiled}.



\subsection{Communication between local and remote MultiCam modules}
\label{sec:comm-betw-local-and-remote}

MultiCam instances on different machines communicate with each other
via the so-called \emph{Skype Desktop API}~\cite{SkypeAPI2011}.  This
is a publicly-documented protocol which third-party applications can
use to communicate with Skype instances. The protocol features dozens
of message types.  To give just two simple examples, the protocol
enables an application to determine the user name of the currently
logged in Skype user, and to be notified whenever a Skype call begins
or ends.

Of particular interest for MultiCam is the set of message types
designated as ``application to application'' (AP2AP) messages.  These
message types enable a third-party application communicating with one
instance of Skype to exchange arbitrary strings of bytes with an
application communicating with another instance of Skype on a remote
computer.  We employ this AP2AP feature to enable two instances of
MultiCam to communicate.  In particular, a local instance of MultiCam
can instruct a remote instance to advance its camera setting, during a
Skype call.  As Figure~\ref{fig:communication} shows, the path
taken by such a request is rather circuitous, requiring four hops
between the various local and remote modules, and employing three
different protocols en route. The precise set of AP2AP message types
employed for communication between MultiCam instances is described in
Appendix~\ref{sec:MultiCam-Ap2Ap}.


\subsection{Implementation of camera-switching}
\label{sec:switching-implementation}

The choice of mechanism for switching cameras is perhaps the chief
design decision for a multi-camera virtual camera.  Let us initially
ignore the possibility of tiled mode and concentrate on switching the
current primary camera between two or more physical cameras.  There
are at least two obvious alternatives, which we will call
\emph{one-at-a-time} and \emph{all-at-once}.  

The one-at-a-time approach uses exactly one camera at any instant:
the DirectShow graph consists of the current primary camera as a
source filter, connecting to the virtual camera filter, which
probably does nothing but pass the physical camera's frames
untouched to the downstream filter.  In this approach, the software
needs to perform surgery on the DirectShow graph whenever the
input camera is switched. Specifically, the software performs a
\texttt{stop} operation on the graph, replaces the source filter
with a filter whose source is the newly-requested camera, then
performs a \texttt{start} operation on the graph.  As we will see
in Section~\ref{sec:experiment3-camera-switching-latency}, there
is evidence this can impose additional latency of 400--700~ms,
compared with the all-at-once approach described next.  


The all-at-once approach connects source filters from all desired
physical cameras to the virtual camera filter when the DirectShow
graph is first created.  Video data is continuously streamed from all
cameras simultaneously, and the job of the virtual camera filter is to
pass on the frames of the current primary camera while dropping data
from the other cameras on the floor.  Clearly, this is extremely
wasteful in terms of computational resources.  However, it has the
benefit of rapid camera switching, as the costly graph surgery
operation is eliminated (see Figure~\ref{fig:camera-switch-latency} in
Section~\ref{sec:experiment3-camera-switching-latency}).  Note that
the all-at-once approach also permits arbitrary combinations of the
input images, such as a tiled view of all cameras, or small overlay
views of the other cameras placed on top of the primary camera view.


Perhaps it is possible to implement a hybrid approach that retains
most of the benefits of all-at-once and one-at-a-time.  For example,
it may be possible to construct a graph in which source filters for
all cameras are present and connected, and the cameras are activated
and ready to transmit data.  But we arrange for data to be transmitted
only on request, so that only one camera is streaming data to the
virtual camera filter at any one time. It is conceivable that such an
approach would consume little or no more resources than the
one-at-a-time approach, while requiring only a small additional
latency for switching.  Investigation of this possibility is left for
future work.  Do cameras even support the notion of being ready to
transmit, without actually transmitting?  If not, this is a feature
that camera manufacturers and developers of frameworks such as
DirectShow might consider implementing in future.  Yet another
possibility is for the non-primary cameras to continuously transmit,
but at a very low resolution suitable for thumbnail images. This would
consume fewer resources while still providing everything the MultiCam
filter needs while in non-tiled mode.  But switching might be slow
with this approach, since multiple cameras would need to change
resolution on every switch.

Note that, in principle, both the all-at-once and one-at-a-time
approaches permit the virtual camera filter to be a so-called
\emph{in-place} transform.  This means that the video data is not
copied: it remains in a buffer and a pointer to this buffer is passed
to the downstream filter.  Obviously, the advantage of an in-place
transform, compared to a more general transform that copies the data
into a separate buffer, is that it avoids the computational expense of
copying.  The filter could even add overlay windows (e.g.\ small views
from other cameras) without sacrificing the in-place property.
However, the MultiCam design does not use an in-place transform.
There are two reasons for this. First, in-place transforms are not
appropriate for tiled mode (discussed next).  Second, if we are using
the all-at-once approach, and the cameras do not all support exactly
the same resolution, it is impossible to switch between cameras
without performing DirectShow graph surgery (see
Section~\ref{sec:resolution}).

The design choices become murkier when we wish to support tiled mode,
or other similar notions such as picture-in-picture.  If we are
displaying video data from all cameras simultaneously, we must take
the all-at-once approach.  Moreover, as mentioned above, tiling the
frames requires that they are subsampled and moved to the correct
subrectangle in the downstream image, which removes the possibility of
using an in-place transform.  One can imagine designs in which the
system switches between all-at-once (or at least some-at-once) when
multiple inputs are required, but falls back to one-at-a-time at other
times. It is not clear that the complexity of the solution would be
justified by its benefits.

Motivated chiefly by the goal of low-latency camera-switching, but
also the desire for a tiled mode, MultiCam uses the all-at-once
approach.  Indirect evidence (especially the switching latency
measured in
Section~\ref{sec:experiment3-camera-switching-latency}) suggests
that the other virtual camera tools analyzed in this report all use
the one-at-a-time approach.  This is not surprising: as discussed
in Section~\ref{sec:related-work}, low-latency camera-switching is
not a primary goal of these tools.

Thus, we now understand the reasoning behind the following two
important design decisions: (i) MultiCam copies data rather than using
an in-place transform, and (ii) MultiCam uses the all-at-once approach
for switching between cameras.

\subsection{Managing heterogeneous resolutions, formats and frame
  rates}
\label{sec:resolution}

What is the resolution of a virtual camera filter? A simple answer is:
the same resolution as the physical camera currently being used as
input.  This makes good sense in the one-at-a-time model, but is
problematic for the all-at-once approach, where the set of cameras may
be heterogeneous and offer different resolutions and aspect ratios.

A brief discussion of webcam capabilities is needed before we
continue.  For our purposes, a \emph{capability} of a webcam is a
tuple expressing a resolution and aspect ratio (e.g.\ $640\times480$),
a compression scheme (e.g.\ none, or H.264), a color format (e.g. RGB
or YUV), and a frame rate (e.g. 30~fps).  Webcams typically offer many
capabilities (one or two dozen is not uncommon).  Software employing a
webcam can designate which of the available capabilities should be
used before any video data is obtained.  Only one capability at a time
is active.

Hence, in the one-at-a-time model, the virtual camera software can
enumerate the capabilities of the (unique) primary camera, select
a capability whose resolution is closest to some target resolution
(possibly using additional criteria such as the compression
scheme), and plug the resulting source filter directly into the
DirectShow graph.  Because the graph is stopped before changing
source filters, any change in resolution---or the remaining
capability dimensions---is handled gracefully.

The all-at-once model does not have this luxury.  The graph is not
stopped during a camera-switch, so it is preferable that the virtual
camera filter maintains a fixed output resolution, even when its input
resolution changes.  MultiCam uses a simple approach to deal with
these problems. It has a fixed (but configurable) target image height,
which is 640~pixels in all experiments reported in this report.  At
startup, each camera's capabilities are enumerated and a capability
with the largest height that fits within the target height is
selected.  MultiCam requests uncompressed RGB data at 30~fps if it is
available.  The motivation for this is that we need to choose some
fixed output format for the MultiCam filter, and uncompressed, 30-fps
RGB is a simple, uncontroversial choice likely to be supported by many
webcams.  If this format is not available, MultiCam requests the first
available alternative at the previously-selected resolution.  In this
case, DirectShow automatically inserts a suitable transform filter
between the source and MultiCam filter, which increases overhead but
is not otherwise problematic.  When preparing a camera frame for
output, if the frame does not fill the target resolution, MultiCam
shifts it appropriately so the output is centered and has a black
border.  A similar process occurs for creating a tiled view, but
images are also subsampled to fit within a designated tile.

MultiCam takes no explicit steps to address issues of timing and frame
rate. It relies on default DirectShow behavior to manage the flow of
data within the graph, which may include dropping frames if a given
filter is operating faster than a downstream filter.

\subsection{Other design and implementation details}
\label{sec:design-details}

MultiCam incorporates or adapts several modules of code not written by
the author.  First and foremost, MultiCam relies heavily on the
Microsoft example code that ships with DirectShow.  Of particular note
here are the generic source filter, transform filter, and
transform-in-place filter examples.  Techniques specific to virtual
cameras were adapted from the publicly-available sample code known as
``vcam.''\footnote{The code is available as
  \url{http://tmhare.mvps.org/downloads/vcam.zip}, and is attributed
  to a person whose name is given only as ``Vivek.''}  MultiCam's
techniques for capturing global mouse and keyboard events rely on an
online article and code by George
Mamaladze~\cite{mamaladze04}\footnote{Code available from
  \url{http://globalmousekeyhook.codeplex.com}.}.  Finally, MultiCam
interacts with the Skype Desktop API via an altered version of a
library written by Gabriel Szabo~\cite{szabo06}.

The MultiCam filter is implemented in approximately 9000~lines of \cpp;
the MultiCam application is implemented in approximately 4000~lines of
C\#; the grand total is therefore about 13,000~lines of code for the
entire package.  (Lines of code are measured by the UNIX utility
\texttt{wc}, which includes comments and blank lines in the count.)
As remarked above, much of the MultiCam code consists of imported
libraries or example code, some of which was substantially altered. A
rough estimate suggests that just under half of the MultiCam code was
imported with little or no alteration. This implies the total amount
of code written, or substantially edited, for MultiCam is
approximately 6500 lines of code.

Additional design details are discussed in Appendix~\ref{sec:mult-design-deta}.




\section{Experience with MultiCam}
\label{sec:typical-usage}

At the time of writing, MultiCam has been employed for a genuine Skype
chat approximately once per week by the author, over a period of five
months.  Here, ``genuine'' means that the chat was not part of a
deliberate experiment, and its primary purpose was communication with
friends or family.  In every case, the reason for using multiple
cameras was that one or more additional family members were present
and I wanted to include them in the video stream.  Obviously, the
impressions gained from this experience have little scientific rigor:
there was no methodical data collection, and my subjective impressions
are probably biased by a desire for MultiCam to appear useful.  The
results of a careful user study are reported in
Section~\ref{sec:user-study}, but that study focuses on one very
specific scenario, whereas my own use of MultiCam has been much more
varied. Therefore, this Section reports briefly on some aspects of my
experience.

With rare exceptions, the remote participants showed little interest
in controlling the cameras. In general, therefore, I was not relieved
of the burden of camera-switching.  On the other hand, I felt the
total effort of camera management was significantly reduced in most
cases.  Rather than constantly having to adjust a single camera to
show the current region of interest, I was frequently able to leave
the cameras in a fixed position for long periods and simply switch
between them.  My enjoyment of the conversations was thus increased.
Remote participants also gave a strong impression of having increased
enjoyment, compared to single-camera conversations in the past.  (But
recall that these participants comprised friends and family, so their
reactions probably have a positive bias.)

Figure~\ref{fig:usage-scenarios-intro}(a)--(c) shows the three camera
setups that proved most useful in these conversations.  In
Figure~\ref{fig:usage-scenarios-intro}(a) we see a two-camera scenario
in which one camera is perched on a laptop for a headshot of the main
Skyper, and another camera is on the table beside it trained on a
child in the background. Figure~\ref{fig:usage-scenarios-intro}(b)
shows another two-camera scenario, again with one camera capturing the
standard Skyper headshot.  The other camera is also perched on the
laptop, but faces the opposite direction. This mimics the setup of
dual-camera smartphones and tablets, but with more flexibility, since
the exact direction of the cameras can be adjusted individually.  In
this scenario, I often pick up the outward-facing camera and direct it
manually for a period of time before placing it back on the laptop.


Figure~\ref{fig:usage-scenarios-intro}(c) shows a three-camera
scenario. Skype is still being run from a laptop, but using a living
room TV as a display.  The remote participant's tiled mode view of
this scenario is shown in Figure~\ref{fig:usage-scenarios-intro}(e).
One camera is mounted on top of the TV, showing a wide view of the
entire scene. Another camera is perched as usual on the laptop for a
headshot of the main Skyper.  A third camera is available to be moved
around as needed, capturing the activity of a small child on the
floor; at this particular instant, the third camera is behind the
Skyper on the arm of a sofa.  This setup has been particularly
successful for group events, such as opening presents, in which
attention naturally focuses on different people at different times.


\section{User study}
\label{sec:user-study}

A user study was conducted to examine some of the benefits and
drawbacks of using multiple cameras with video chat, focusing
especially on a comparison between speaker-controlled and
listener-controlled camera-switching.  

\subsection{Participants}

A group of 23 individuals was recruited to participate in the
study. Participants were all acquaintances of the author who
voluntarily responded to email requests or similar; the resulting
participant pool comprised friends, family, colleagues, and one
student.  Participants' ages ranged from 20 to 70 (median 40).  Two
participants were new to Skype; the remainder had frequently used
Skype for single-camera video chat. Two participants had used the
MultiCam camera-switching functionality previously; of the remainder,
four had some knowledge of the MultiCam project, and the remaining 17
participants had no knowledge of it.  Nine of the participants could
reasonably be described as technically savvy (i.e.\ work in a
computer-related profession, or maintain an active amateur interest in
technology); the remainder had no particular skills or affinity with
computer technology.  Geographically, there was a three-way split
between participants: five in the same North American town as the
author, eight in other North American locations, and ten outside North
America (all either Europe or Oceania).  Approximately 70\% of
participants employed laptop monitors, with the remainder using larger
desktop monitors. Fourteen users employed a single webcam at their own
end of the conversation; nine used no camera at all; none used
multiple cameras.  Hence, although the sample is relatively small and
was not selected via random sampling, it contains a good cross-section
of video chat users.

\subsection{Method}

The user study employed the two-camera setup shown in
Figure~\ref{fig:usage-scenarios-intro}(d), in which a person (the
speaker) can sit on a sofa and communicate with the study
participant (the listener), using a whiteboard adjacent to the sofa
when desired.  We will refer to this video chat scenario as the
\emph{whiteboard lecture scenario}.  One camera, positioned on top of
the laptop, presents a head-and-shoulders view of the speaker
sitting on the sofa.  The other camera, positioned on the desk,
displays the whiteboard.  Thus, exactly 3 views were available to
study participants: the speaker, or the whiteboard, or a tiled
view of both.  (The tiled view is shown in
Figure~\ref{fig:usage-scenarios-intro}(f).)  The whiteboard is
positioned such that, on a typical monitor and under typical Skyping
video quality, writing on the whiteboard can be read reasonably easily
when the whiteboard camera is the primary camera, but is not very
legible in the tiled view.  This is important because it provides an
incentive to switch between views; otherwise, it would probably be
optimal to remain in tiled view at all times, and this would reveal no
useful information comparing local and remote camera control.


As will be described in more detail shortly, participants needed the
ability to switch between the three camera views in this study. As
explained in Section~\ref{sec:MultiCam-overview}, the only
camera-switching method guaranteed to be available to all users is the
switch-by-IM method.  For consistency, therefore, all participants
used the IM method for switching cameras in this study.

Each user in the study participated in a Skype session with the
author, lasting about 10 minutes.  The core of the session involved
two three-minute lectures, delivered by the author using the
whiteboard and a handheld prop.  
The most important feature of the session was that in one of the
three-minute lectures, the speaker had exclusive control of the
camera-switching, and in the other lecture, the listener had exclusive
control.  The ordering all these two camera-control options was
alternated for each participant, so the first lecture was
speaker-controlled in exactly half of the sessions.

The following script describes the content of each session in more detail:
\begin{enumerate}
\item Introduce the physical environment, and demonstrate the three
  possible views (person, whiteboard, tiled).
\item Explain to the listener how to switch between the three views
  using Skype instant messages, and allow the listener sufficient
  practice at switching views until they claim to be comfortable with
  it (typically 30--60~seconds).
\item Explain that the listener will now receive two three-minute
  mini-lectures about a particular topic in computer science (data
  compression), and the speaker will control the cameras in one
  lecture, whereas the listener will control the cameras in the other
  lecture.  Announce who will be controlling the cameras first. (As
  already mentioned, this alternated with each participant.)
\item Deliver first lecture, on run-length encoding.
\item State clearly that responsibility for camera control is now
  switching from speaker to listener or vice versa.
\item Deliver second lecture, on LZ77 compression.
\item Verbally administer the questionnaire.
\end{enumerate}

Both mini-lectures involved the same routine of alternately talking
directly at the camera while sitting on the sofa, and writing on the
whiteboard.  The specific set of states for each lecture was: sofa,
whiteboard, sofa, whiteboard, sofa.  The middle ``sofa'' segment
involved, for both mini-lectures, the use of a handheld prop (actually
a paperback book that was opened to show some example data that we
might wish to compress).  Hence, even listeners who might have been
happy to stare at a whiteboard while listening to a disembodied voice
had an incentive to switch back to the sofa view during the middle
segment.

Precise details of the questionnaire administered at the end of each
session are given in Appendix~\ref{sec:user-study-questionnaire}.  The
most important questions gauged whether the users preferred
speaker-controlled cameras, listener-controlled cameras, or neither.
The strength of this preference was coded using Likert-type categories
(e.g.\ ``mildly disagree'').  Other questions asked users to list any
aspects of the experience they liked or disliked during the
speaker-controlled and listener-controlled segments.  Users were also
asked how much they used the tiled view, and a final open-ended
question asked for any further comments or feelings about the
experience.

It is important to note that it is definitely not the goal of the
study to evaluate the raw efficacy of the whiteboard lecture scenario
for distance learning or collaborative web conferencing.  The scenario
is contrived solely to provide an easily-controlled, replicable
situation in which remote and local control of camera-switching can be
compared while keeping other factors constant.  Indeed, numerous
software products targeted at distance learning and web conferencing
are available,\footnote{e.g.\ Elluminate and Wimba Classroom for
  distance learning; GoToMeeting and Microsoft's Live Meeting for web
  conferencing---to mention just two of the many products available in
  each category.} and MultiCam is not envisaged as a direct competitor
these products.  In fact, they are complementary: any such product
receives input from a webcam, and can therefore be enhanced by using
MultiCam-style virtual camera software to provide simultaneous
multiple-camera functionality if desired.

\subsection{Results and discussion of user study}
\label{sec:user-study-results}

\subsubsection{Camera control preference}

\begin{figure}[htb]
  \centering
    \renewcommand{\thiswidth}{140mm}
    \includegraphics[width=\thiswidth]{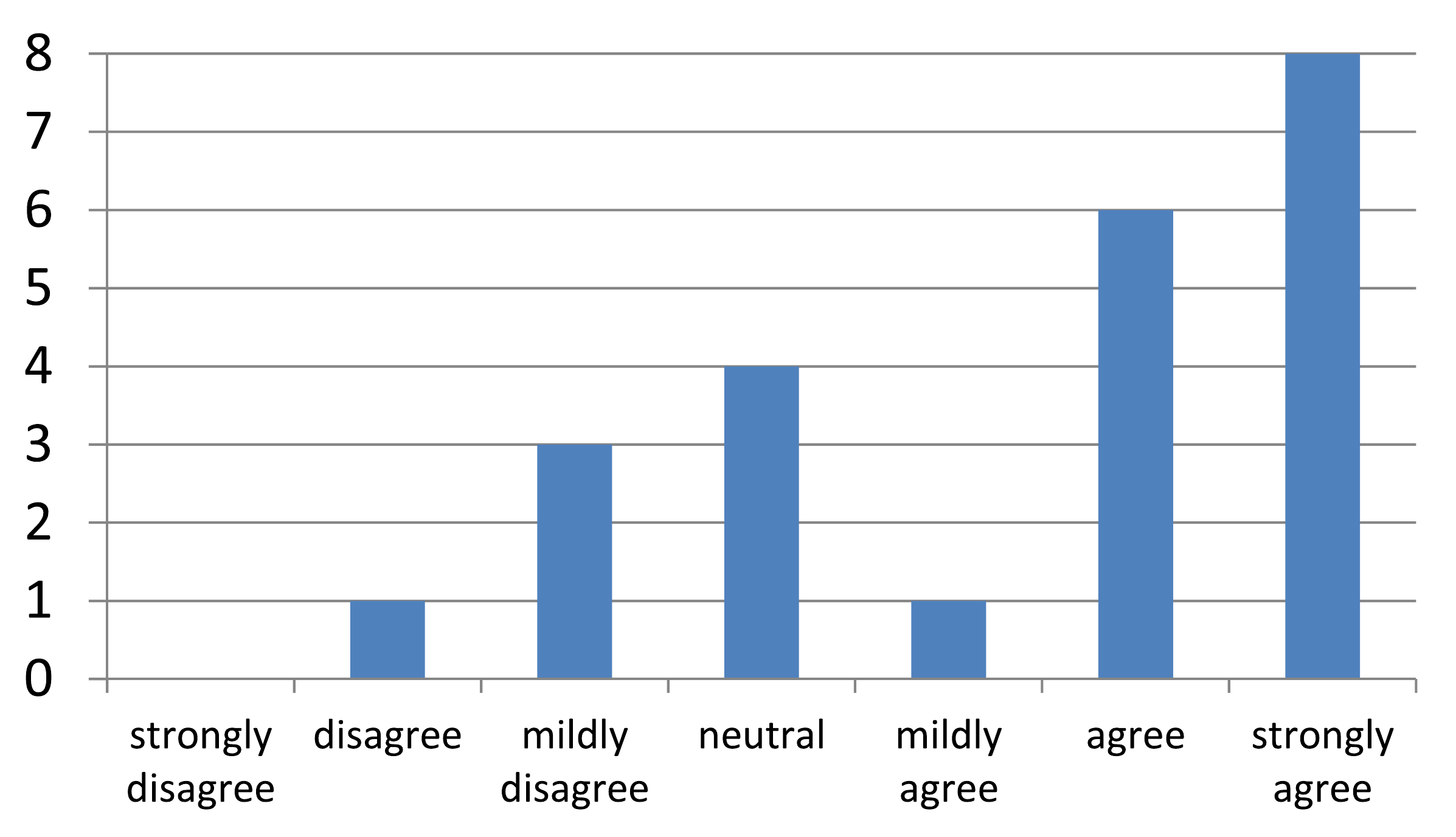}
    \caption{\textbf{User preferences for speaker-controlled
        camera-switching vs listener-controlled camera-switching.}
      Frequencies of agreement levels with the following statement are
      shown: ``When the speaker controlled the camera, the overall
      experience was more satisfactory.'' }
  \label{fig:user-study-agreement}
\end{figure}

Figure~\ref{fig:user-study-agreement} shows the strength of
participants' preferences between speaker-controlled and
listener-controlled camera-switching. For simplicity, the graph shows
results coded with Likert-type categories (i.e.\ the level of
agreement/disagreement) applied to the statement ``When the speaker
controlled the camera, the overall experience was more satisfactory.''
However, to eliminate acquiescence bias,\footnote{\emph{Acquiescence
    bias} is the tendency of respondents to agree with statements.
  See texts on psychology or market research for details
  (e.g.~\cite{richards08:psych}).}  the data was obtained in a
different way, resulting in perfect symmetry between preferences for
listener control and speaker control.  Participants were first asked
whether they preferred speaker control, listener control, or neither.
Those expressing a preference were then asked to follow up by
selecting from ``strongly agree,'' ``agree,'' or ``mildly agree'' in
reaction to the statement ``When the [speaker/listener] controlled the
camera, the overall experience was more satisfactory.''  Of course,
the word ``speaker'' or ``listener'' in this statement was selected
according to the participant's previously-stated preference.

A glance at Figure~\ref{fig:user-study-agreement} gives the strong
impression that users preferred speaker-controlled camera-switching,
and this impression is confirmed by statistical analysis. The median
response is ``agree''---the second-highest response on the 7-point
scale.  To check our intuition that this median differs by a
statistically significant amount from the ``neutral'' response, we can
perform a chi-squared test of the null hypothesis that the population
median is ``neutral.''  To do this, restrict attention to the
19~participants who expressed a preference: 4~for listener control and
15~for speaker control. If the null hypothesis held, we would expect
$9.5$~in each category. Computing a chi-squared statistic in the usual
way, we find $\chi^2=6.37$ on 1~degree of freedom, which yields a
$p$-value of 0.012.  Hence, we can reject the null hypothesis at, say,
the 2\% level of significance, and conclude that there was a
statistically significant preference for speaker control.

On the other hand, we also see that the results were not a complete
landslide for speaker-controlled camera-switching: 15~participants
expressed a preference for speaker control, and 8~did not. So 35\% of
respondents are either neutral or prefer listener control.  Applying
the usual formula for standard error $\sigma$ of a population
proportion gives $\sigma=(0.35\times(1-0.35))/\sqrt{23}=0.052$.  The
95\% confidence interval of $\pm2\sigma$ is therefore $[24\%, 45\%]$.
In fact, the sample size is too small for this simple approach to be
rigorous, but it seems adequate for the qualitative discussion given
here.  We only wish to conclude that a significant minority (perhaps a
quarter to a half) of the population does \emph{not} prefer speaker
control.


Combining the conclusions of the previous two paragraphs, we see that
for the particular whiteboard lecture scenario tested, an ideal
multi-camera system would function primarily by speaker-controlled
switching, to satisfy the statistically-significant preference of the
population for speaker control. However, the ideal system would also
permit control by the listener (to whatever extent desired), which is
especially important for the significant minority of listeners who
prefer to be in control. Obviously, we should be extremely careful
when extrapolating this conclusion beyond the particular version of
the whiteboard lecture scenario tested.  We expand on this point in
Section~\ref{sec:gener-concl-from}, after presenting the remaining
user study results.

\subsubsection{Pros and cons of camera-switching options}

Figure~\ref{fig:theme-analysis} lists all the important themes to
emerge from the questions asking participants to list any likes or
dislikes of the two camera-switching options (speaker control and
listener control).  Note that these were open-ended
questions,\footnote{Advantages and disadvantages of open-ended
  questions are discussed in psychology and market research textbooks
  (e.g.~\cite{ray11:_method_towar_scien_behav_exper}).} so
participants had no cues as to possible responses. The Figure shows
any theme that was mentioned by at least two participants.
Classification of responses was done by the author, and is of course
subjective.  Nevertheless, several clear points emerge.

\begin{figure}[tb]
  
  \renewcommand{\thiswidth}{80mm}
  \renewcommand{\thiswidthB}{4mm}

  Advantages of speaker control:

    \begin{center}
      \begin{tabular}{|m{\thiswidth}|m{\thiswidthB}|}
        \hline 
         could concentrate more easily 
         (not distracted by thinking about switching cameras) &  $\,\,$9 \\
        \hline 
         lecturer can anticipate the need for a switch and 
         thus switches at the right time                      &  $\,\,$8 \\
        \hline 
      \end{tabular}
    \end{center}

  Advantages of listener control:

    \begin{center}
      \begin{tabular}{|m{\thiswidth}|m{\thiswidthB}|}
        \hline 
        had control over the experience                       & 10 \\
        \hline 
        had the ability to go back to the whiteboard when 
        desired                                               & $\,\,$5 \\
        \hline 
        concentrated better because had to pay attention      & $\,\,$2 \\
        \hline 
      \end{tabular}
    \end{center}

  Disadvantages of listener control:

    \begin{center}
      \begin{tabular}{|m{\thiswidth}|m{\thiswidthB}|}
        \hline 
        poor interface for switching cameras & $\,\,$8 \\
        \hline 
        harder to concentrate/distracting to switch cameras   &  $\,\,$5 \\
        \hline 
        switching delay was annoying & $\,\,$4 \\
        \hline 
        lose a few seconds of attention at every switch & $\,\,$3 \\
        \hline 
      \end{tabular}
    \end{center}

    \caption{\textbf{Theme analysis of user study comments.}  All
      comments that occurred twice or more are listed, with the
      frequency of occurrence in the right column.}
  \label{fig:theme-analysis}
\end{figure}

The strongest reason for liking speaker control was that it was easier
to concentrate on the content of the lecture---these participants
considered camera control a burden, and devoting thought to camera
control detracted from the attention that could be paid to the lecture
itself.  For example, one participant stated: ``I can concentrate on
the speaker, not on the technology.''  A related but separate point is
that the speaker knows in advance when a switch will be required, and
thus is able to time the switches appropriately.  In contrast, the
listener realizes a switch is required only \emph{after} the speaker
performs whatever action triggers the need for a switch.  Thus, even
for a user who does not find camera-switching burdensome, listener
control has the disadvantage that most camera switches occur late. One
participant spoke of losing ``a few seconds'' of relevant viewing at
every such switch.

No themes for disliking speaker control emerged; the only comment in
this category was from a single participant, who noted that he or she
``couldn't check something on the whiteboard.''

The strongest reason for liking listener control was the somewhat
tautological notion of being ``in control.''  Some participants
perceived explicit educational value in being able to time their own
switches, especially for lingering on, or extra glances at, the
whiteboard.  In fact, four of the five users who mentioned the ability
to go back to the whiteboard as an advantage of listener control
actually preferred speaker control in general.  This is important, as
it demonstrates that even users who prefer speaker control can benefit
from the ability to seize control occasionally.  A more subtle and
surprising effect was also apparent: some users derive intrinsic
satisfaction from being in control, without necessarily perceiving a
causal link to an educational outcome. Comments along these lines
include: ``it was kind of fun to be the one in charge,'' and ``the
part of me that likes to flip through the channels liked it.''  Two
participants preferred listener control for another surprising reason:
they found the requirement to be alert and ready to switch cameras
when necessary forced them to pay more attention to the lecture,
resulting in a more satisfactory outcome.  This reasoning directly
contradicts the 10 users who found camera-control detrimental to
concentration---more evidence that the user base has diverse
preferences and multi-camera video chat should try to account for
them.

The main stated disadvantage of listener control was the poor
interface for switching cameras. There were two aspects to this. As
remarked above, remote camera-switching was performed via the IM hack,
which requires a minimum of two keystrokes and, more importantly, is
not at all intuitive.  It is not surprising that users disliked this.
However, six users were also frustrated by having to cycle through
the three view settings in a fixed order.  This calls into question
one of the hypotheses on which the MultiCam interface was based:
namely, that switching between multiple views, including a tiled view,
is excessively complex and that the simplest possible interface (a
single advance-to-next-view operation) is therefore preferable.  It
seems this hypothesis is not correct for a significant fraction of
users. Thus, alternative interfaces should be explored in future
multi-camera chat systems.

Another important dislike of listener control was the delay between
requesting a switch and receiving it. Average round-trip times were
not recorded during the user study chat sessions, so it is not known
if these complaints correlate with large network latencies.  (Two of
the four who mentioned this problem were in Oceania, but the other two
were in North America---the same continent as the lecturer.)  In any
case, it is interesting that delay was perceived as a disadvantage
specific to \emph{listener} control.  Speaker-initiated switches would
have suffered delays of similar magnitude (although perhaps up to 50\%
less, depending on the root cause), but were not perceived as
problematic.

There is one other factor that was notable for its absence in the user
comments: when the listener controls the cameras, the speaker is freed
from this concern, and (at least in principle) can devote more effort
to delivering a smooth and clear presentation.  Of course, in this
case, the lecturer was extremely well-practiced in using MultiCam and
was able to use the convenient camera-switching interface provided by
the MultiCam application.  So it is quite possible there was little or
no discernible difference between the clarity of the two
mini-lectures.

\subsubsection{Use of tiled mode}

It is natural to wonder whether multi-camera video chat systems should
provide a tiled mode: is it a beneficial feature, or does it just
clutter the interface and confuse the users? The user study was not
specifically designed to answer this question, and the utility of
tiled mode clearly depends on the application.  Nevertheless, we can
glean a little insight from the participants' responses.  Two
participants chose to use tiled mode most of the time during the
listener-controlled mini-lecture. A further nine participants used
tiled mode at least once.  The remaining 12 participants did not use
tiled mode.  Hence, it seems that for this application at least, tiled
mode is attractive to a significant fraction of users.

\subsection{Conclusions from the user study}
\label{sec:gener-concl-from}

The nutshell conclusion of the user study is: for the
whiteboard lecture scenario, a majority of users prefer
speaker-controlled camera-switching to listener-control, but a
significant minority do not.  Note, however, that care is needed when
extrapolating this conclusion beyond the particular version of the
whiteboard lecture scenario tested.  Indeed, even if we restrict
consideration to the whiteboard lecture scenario, it seems clear
that generalization is problematic.  This is because certain aspects
of the scenario could be varied in such a way as to produce
preferences tilted strongly towards speaker or listener control.  For
example, the speaker could have deliberately ``forgotten'' to switch
cameras several times during the speaker-controlled test.\footnote{In
  fact, this did happen twice, by accident.  Participants were
  instructed to disregard the mistakes, but they may have been
  influenced anyway, of course.} This would be immensely frustrating
to the listeners, and could be made as extreme as desired, resulting
in virtually 100\% of participants expressing a preference for
listener control.  On the other hand, the speaker could have made
listener control difficult and frustrating by frequently moving on and
off the whiteboard, picking up props for only one or two seconds, and
making very brief references back to the whiteboard, all without
verbally telegraphing any intentions.

These thought experiments demonstrate that preference for listener- or
speaker-control is highly application-dependent.  And there are two
other factors that may have influenced the results: (i) the use of the
non-intuitive IM hack for switching cameras; and (ii) the fact that
the vast majority of participants had never used MultiCam before, and
had only a brief 30--60-second practice session to gain familiarity
with switching cameras. Both of these factors would tilt the results
towards a preference for speaker control.

But the application-dependence and other sources of variability do not
render our conclusions from the user study irrelevant---they simply
mean we must be careful in making generalizations.  For example, it
would be wrong to conclude that a majority of users prefer
speaker-control to listener-control for multi-camera video chat in
general.  On the other hand, it does seem reasonable to infer the
following conclusions:
\begin{itemize}
\item For any given multi-camera video chat scenario, there can be
  both a significant proportion of users who prefer local control of
  camera-switching, and a significant proportion of users who prefer
  remote control.
\item Even users who have a preference for not controlling the
  camera-switching in a given scenario can derive benefits from
  seizing control occasionally.
\item A significant fraction of unpracticed users find that
  controlling the cameras detracts from their ability to concentrate
  on the video chat (but this may not be true of users with
  substantial practice, especially if a more convenient interface than
  the IM hack were provided).
\item Significant delays between a switch request and its execution
  can be a source of frustration.
\end{itemize}


\section{Benchmark experiments}
\label{sec:benchmarks}

If, as this report suggests, end-users can benefit from the use of
multiple cameras while video chatting, it is important to verify that
simultaneous use of multiple cameras does not consume excessive
resources on a consumer-grade computer.  This section describes an
experiment to investigate this, together with additional experiments
that examine some MultiCam design choices and compare the performance
of MultiCam to two other multi-camera systems.


\subsection{Hardware used by the experiments}
\label{sec:hardw-used-exper}

The experiments employ four different USB webcams: a Logitech QuickCam
Chat, a Logitech QuickCam Easy/Cool\footnote{The Logitech QuickCam
  comes in several different flavors, which are not always clearly
  distinguished by vendors. The names given here are the ones the
  cameras themselves advertise to DirectShow as a so-called
  \emph{friendly name}.  In practice, this is the string that user
  sees when selecting a camera source within a video chat program.}, a
Microsoft LifeCam VX-3000, and a Microsoft LifeCam HD-3000.  These are
all low-cost cameras: at the time of writing, they could be purchased
from Amazon at costs of \$19, \$21, \$25, and \$29 respectively.  We
deliberately use low-cost cameras, as we are targeting consumers who
have no desire to purchase professional-grade equipment. The selection
of cameras is heterogeneous for two reasons: (i)~it allows us to
investigate the amount of variability in resource usage and
performance between these cameras, and (ii)~it is perhaps more
representative of a consumer whose collection of webcams has grown
piecemeal over time.

Experiments were conducted on two different machines: a relatively
recent (2011) standard office desktop with four cores, and an older
(2007) laptop with two cores.  The motivation behind the choice of
machines is that the desktop represents a typical consumer setup at
the time of writing, whereas the laptop could be considered an
impoverished setup.  If we are prepared to ignore feebler devices such
as smartphones, tablets, and even older single-core machines, our
laptop is a reasonable worst-case scenario.

Figure~\ref{fig:hardware-details} tabulates the some details of the
chosen machines' hardware and software.  Before looking at empirical
results, let us consider the theoretical impact of each of these
specs.  The CPU speed and number of cores have a direct impact on the
CPU utilization measurements, as we will see shortly. A na\"{i}ve
calculation accounting for only the number and clock frequency of the
cores suggests the desktop will be $(4\times 2.66) / (2\times 1.83)
\approx 2.9$ times as fast as the laptop for tasks with plenty of
parallelism.

The amount of memory is irrelevant for these experiments, as long as
there is ample memory to accommodate the maximum of 240~MB consumed by
the most resource-hungry experiment (simultaneously running
MultiCam.exe and Skype.exe during a 4-camera video chat).  The memory
frequency (being directly proportional to the bandwidth between main
memory and CPU) also turns out to be irrelevant: memory bandwidth
proves not to be a bottleneck for either machine, as the following
worst-case calculation shows. A camera operating at 30~frames per
second (fps), with resolution\footnote{The reader may legitimately ask
  why $480\times 640$ can be considered a ``worst-case''
  resolution. Clearly, we can imagine video chats employing much
  higher resolutions than this. Nevertheless, at the time of writing,
  $480\times 640$ is a generous resolution for typical video chat
  scenarios, and appears to be the default Skype resolution for many
  webcams.} $480\times 640$, and transmitting its data in uncompressed
24-bit RGB format, consumes a bandwidth of 26~MB/s. So four cameras
consume about 105~MB/s, which is far less than the 3102~MB/s memory
bandwidth measured on the low-spec laptop, let alone the 6894~MB/s for
the standard desktop machine.

\begin{figure}
  \centering
  \fbox{%
\begin{tabular}{lll}
               & \textbf{desktop} & \textbf{laptop} \\
year purchased & 2011    & 2007   \\
make and model & Dell Optiplex 780 & Toshiba Portege M400 \\
CPU model      & Intel Core2 Quad (Q9400) & Intel Core2 (T5600) \\ 
CPU cores      & 4       & 2 \\
CPU frequency  & 2.66~GHz & 1.83 GHz \\
amount and type of memory & 16 GB dual-channel DDR3 & 4 GB dual-channel DDR2 \\
DRAM frequency & 532 MHz & 333 MHz \\
measured memory bandwidth & 6894 MB/s & 3102 MB/s \\
GPU           & ATI Radeon X1550 & Intel GMA 950 \\
USB 2.0 ports & 8 & 3  \\
Operating system & Windows 7 Enterprise (64-bit) & Windows 7 Professional (32-bit) \\
Skype version & 5.5.59.124 & 5.5.59.124 \\ 
\end{tabular} 
}
\caption{\textbf{Details of the two machines used for resource usage
    experiments.} Memory bandwidth is measured using the STREAM
  benchmark~\cite{McCalpin2007,McCalpin1995}.}
  \label{fig:hardware-details}
\end{figure}

The number of USB ports can also impact experiments of this type. Each
USB 2.0 port has a theoretical maximum bandwidth of 60~MB/s, assuming
each has an independent controller (this assumption is not true, as we
shall see later).  So the worst case camera bandwidth of 26~MB/s is
less than half of the 60~MB/s USB 2.0 bandwidth.  Thus, provided there
are at least as many ports as cameras, there is, in principle, ample
bandwidth for the data from each camera to traverse the USB ports.
Note that the laptop had only three USB ports, and therefore required
two of the cameras to be multiplexed through a USB hub\footnote{A
  D-Link DUB-H7 USB 2.0 powered hub.} for our four-camera experiments.
Again assuming the worst-case camera bandwidth of 26~MB/s, we see
that, in principle, even two-way multiplexing of cameras through a USB
2.0 hub should not produce a bottleneck.  In practice, of course,
there may be more subtle limitations on simultaneously streaming data
from many USB ports.  Some of the variability in the frame rate and
CPU utilization results (look ahead to Figure~\ref{fig:cameras-CPU},
which is discussed in detail below) may arise from this.

\subsection{Experiment 1: MultiCam resource usage}
\label{sec:experiment1-resource-usage}

The objective of the first experiment is to measure the resource usage
of MultiCam with up to four cameras, both in isolation and as part of
a video chat.  Which resources are of interest for this experiment?
Note that network resources are irrelevant, because during a video
chat, the chat software using the MultiCam filter consumes exactly the
same network bandwidth as it would with a single camera, regardless of
the number of physical cameras connected to the MultiCam filter.  In
fact, the two primary resources consumed by the cameras are (i)~CPU,
and (ii)~bandwidth of various internal buses, especially USB buses.
In this experiment we report CPU utilization directly, whereas the
effects of bus saturation are demonstrated indirectly, by measuring
the video frame rate of the MultiCam filter.

There was a separate run of the experiment for each
nonempty subset of the four cameras, resulting in a total of 15~camera
combinations.  As already mentioned, the experiment was repeated on
two different machines (desktop and laptop). Furthermore, each
machine/camera possibility was tested in two different ways: a
\emph{raw run} and a \emph{Skype run} (described in more detail
below).  This results in a total of $15\times 2\times 2=60$ runs.  In
each run, two main data points were collected: CPU utilization, and
video frame rate.  


A \emph{raw run} consisted of executing a simple benchmark program
that displays a MultiCam video stream on the monitor in tiled
mode. Specifically, this was a lightly-altered version of the PlayCap
example code in DirectShow. Note that the raw runs therefore did not
involve any video chat software---the objective was to measure the
bare-bones resource consumption of the cameras connected to the
MultiCam filter, without any additional overhead for video chat.  

A \emph{Skype run} consisted of a Skype video chat between the two
experiment machines described above.  Only one end of the chat
transmitted video in each run, and audio was disabled (although the
impact of this is small).  Let us call the machine transmitting video
the \emph{source} of the chat, and the machine receiving video the
\emph{destination}.  CPU utilization and video frame rate were both
measured at the source, and frame rate was measured via Skype's ``Call
Technical Info'' menu option.  For all Skype runs, the laptop was
connected to a residential broadband service via 802.11g wireless, and
the desktop employed a wired connection to a university campus
network.  The physical distance between the two machines was about
0.5~miles, and Skype reported the vast majority of round-trip times in
the range 50--60~ms. 

In each run, the average CPU utilization was recorded over a period of
90 seconds, using Windows performance counters. CPU utilization
figures are reported as a percentage of all CPU resources on the
machine (e.g.\ 50\% utilization on a four-core machine is equivalent
to 100\% utilization of two of the cores.) Frame rates were averaged
over 250 frames. Naturally, all reasonable steps were taken to prevent
other processes consuming resources during measurement periods. In
addition, each run had a burn-in period of about 10 seconds to allow
the video stream to initialize satisfactorily before measurements
began. As discussed in Section~\ref{sec:design-details}, cameras are
always requested to deliver data at $640\times 480$ resolution with
30~fps\footnote{\label{footnote:frames-per-second}It turns out that
  one of the cameras (the VX-3000) can only support up to 15~fps at
  this resolution.}, in 8-bit RGB format, and are subsequently
subsampled by the MultiCam filter if necessary for display in tiled
mode.



\subsubsection*{Results for raw runs}
\label{sec:results-raw-runs}

\begin{figure}
  \centering
  \begin{tabular}{c}
    \renewcommand{\thiswidth}{100mm}
    \includegraphics[width=\thiswidth]{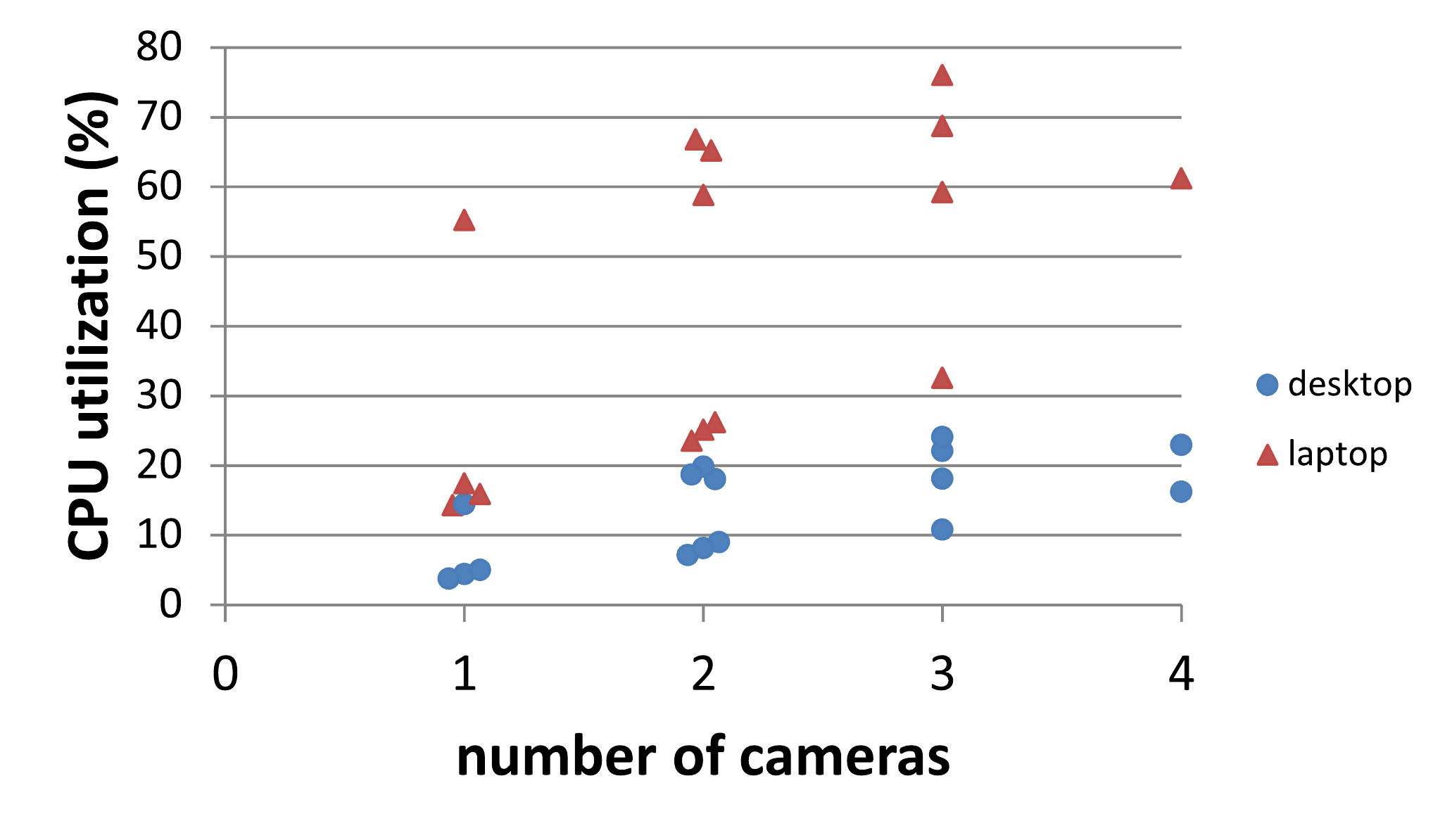} \\ [3mm]
    \includegraphics[width=\thiswidth]{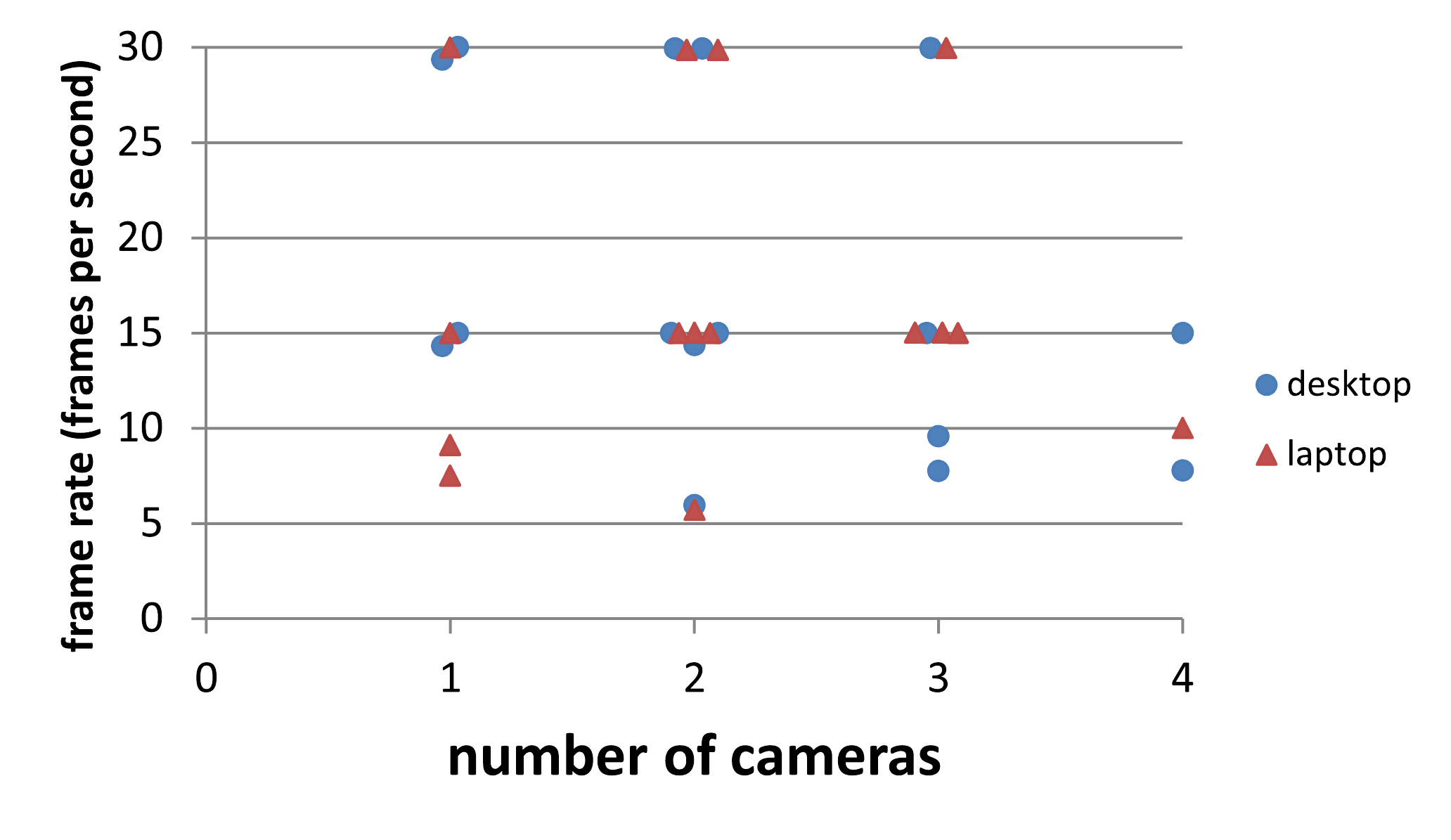} \\ [3mm]
    \includegraphics[width=\thiswidth]{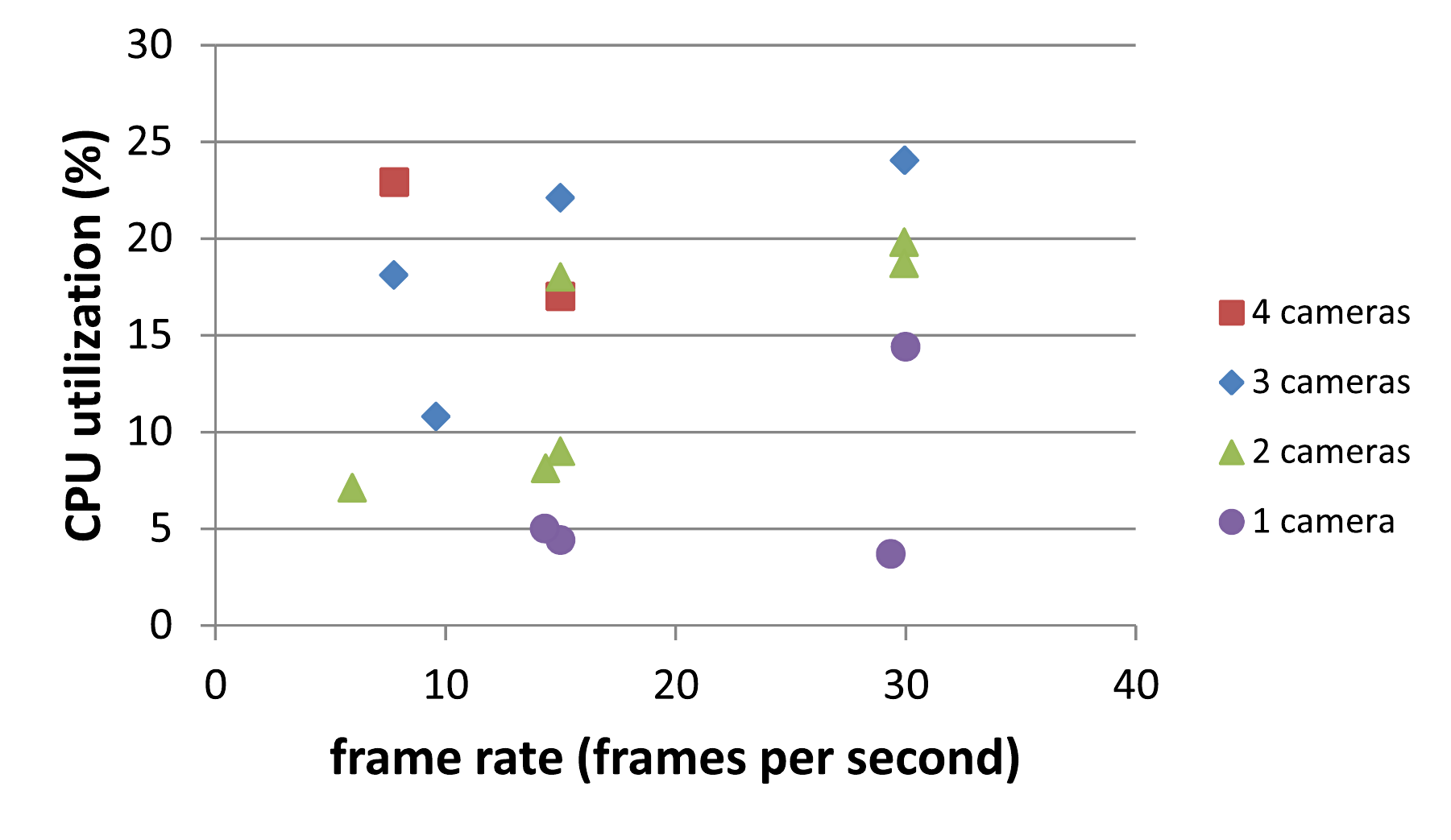} 
  \end{tabular}
  \caption{\textbf{MultiCam CPU utilization and frame rate for a local
      display benchmark.}  Top: CPU utilization for all possible
    combinations of four different webcams, on two different
    computers.  Middle: Frame rate for the same set of experiment
    runs.  Bottom: The same data as the previous two graphs, combined
    a single graph (for clarity, only the runs on the desktop computer
    are shown). In the top and middle panels, some points have been
    shifted horizontally to improve visibility.}
  \label{fig:cameras-CPU}
\end{figure}

Figure~\ref{fig:cameras-CPU} shows the results of the raw runs.  The
top panel shows how CPU utilization varied for the 15 different camera
combinations.  There are several interesting features of these
results. First, the CPU utilization of the laptop is roughly three
times higher\footnote{The exact value of the average multiplier is
  3.3.} than the desktop, as we would expect from the discussion in
Section~\ref{sec:hardw-used-exper} (although there is considerable
variation from the $3\times$ multiplier).  Second, CPU consumption by
the cameras is approximately additive: the cost of any set of cameras
is roughly the sum of the cost of each camera used individually (but
again with considerable variation, and at least one significant
exception discussed below).

Third, there can be great variation in the cost of any given
camera. In this case, three of the cameras consume about 5\% of the
desktop CPU, but the remaining camera consumes about 15\%---three
times as much!  The outlier camera is the LifeCam HD-3000.  No
specific efforts were made to investigate the discrepancy, but it may
be due to the higher native resolution of this camera compared to the
others.  To further complicate matters, it turns out that the HD-3000
consumes only half as much CPU (about 8\%) when plugged into a USB
port on the back of the machine, rather than one of the front ports
that happened to be used for this particular experiment.  This also
explains the two data points in the 4-camera desktop category, in all
three panels of Figure~\ref{fig:cameras-CPU}: in each case, the
HD-3000 was swapped from the front to the back to produce a second
4-camera data point.

This leads to the fourth observation: there can be very significant
variations in resource usage by cameras, and the variation can be for
obscure reasons that would certainly be inexplicable to typical
end-users.  The $2\times$ change by moving from a rear to front USB
port just mentioned is one example of this. Another peculiarity is the
fact that two of the 3-camera sets on the laptop consume significantly
more CPU than the 4-camera set.  This suspicious result was confirmed
in repeated tests, but was not investigated further.


The fifth and final observation is the most obvious but also most
important: the total CPU consumption on our typical (i.e.\ desktop)
setup is only a small fraction of the available resources, and even on
the impoverished (i.e.\ laptop) setup, the most CPU-intensive runs
still leave some room for other tasks to use the CPU. Hence we can
conclude that multi-camera video chat is comfortably feasible on
consumer PC hardware.

The middle panel of Figure~\ref{fig:cameras-CPU} shows the video frame
rate achieved by MultiCam for the same camera sets as the CPU tests.
On this metric, the desktop and laptop have very similar performance
for the majority of camera sets.  But there are notable exceptions,
including the surprisingly low frame rates of two cameras (the two
QuickCams) in single-camera mode on the laptop.  As with the CPU
results, we see some counterintuitive trends. For example, the two
QuickCams, which each achieve fewer than 10~fps when used alone on the
laptop, are even worse when used together (6~fps), but can improve
dramatically when combined with one of the other cameras (e.g.\ 30~fps
with the HD-3000).  Such mysterious results might lead one to suspect
a performance bug in MultiCam, such as misuse or abuse of the
DirectShow APIs.  But as we will see shortly
(Figure~\ref{fig:cameras-CPU-skype}, middle panel), Skype obtains
similarly poor frame rates from one of the QuickCams when using it
directly (i.e.\ without MultiCam) on the laptop.  This suggests that
the poor single-camera performance---and presumably the
mysteriously-good multi-camera performance too---is not specific to
MultiCam, and may instead derive from some combination of the camera
drivers, the USB controllers, and the DirectShow framework.

The bottom panel of Figure~\ref{fig:cameras-CPU} shows the same data
as the two upper panels, combined on a single graph so that any
relationship between frame rate and CPU utilization can be observed.
For clarity, only results for the desktop machine are shown. The
surprising result is that the relationship is rather weak ($R^2=0.27,
0.67, 0.51$ for $1,2,3$ cameras respectively). This once again
demonstrates that camera performance and resource usage is
unpredictable, and appears to depend on subtle interactions between
several hardware and software modules.

The high-level conclusion to be drawn from
Figure~\ref{fig:cameras-CPU} is twofold. First, video chat with
several cameras simultaneously consumes only a fraction of the
resources on a typical consumer machine and is therefore
feasible. Second, the performance (in terms of both CPU and frame
rate) of heterogeneous sets of cameras working together is
unpredictable.  Camera manufacturers and video chat software
developers probably need to devote considerable effort to reducing
this unpredictability if end-users are to experience consistently
satisfying multi-camera video chats.  The good news is that this
experiment did uncover some sweet spots: in the bottom panel of
Figure~\ref{fig:cameras-CPU}, for example, we see that one set of
three cameras can operate at 30~fps for less than 25\%~CPU, and all
four cameras can operate at 15~fps for less than 20\%~CPU, provided
that we are lucky enough to choose the right USB ports.

\subsubsection*{Results for Skype runs}
\label{sec:results-Skype-runs}

\begin{figure}
  \centering
  \begin{tabular}{c}
    \renewcommand{\thiswidth}{100mm}
    \includegraphics[width=\thiswidth]{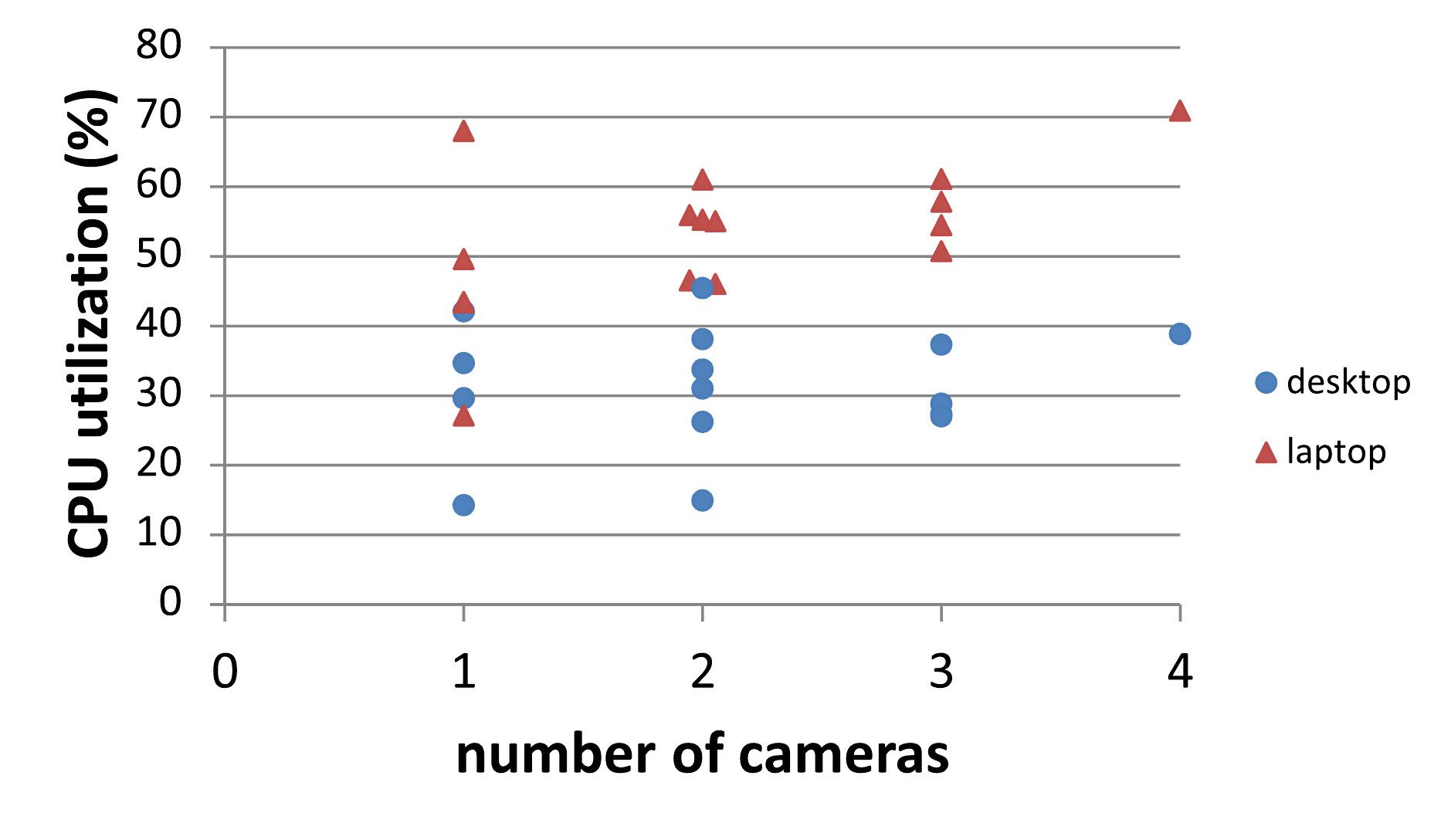} \\ [3mm]
    \includegraphics[width=\thiswidth]{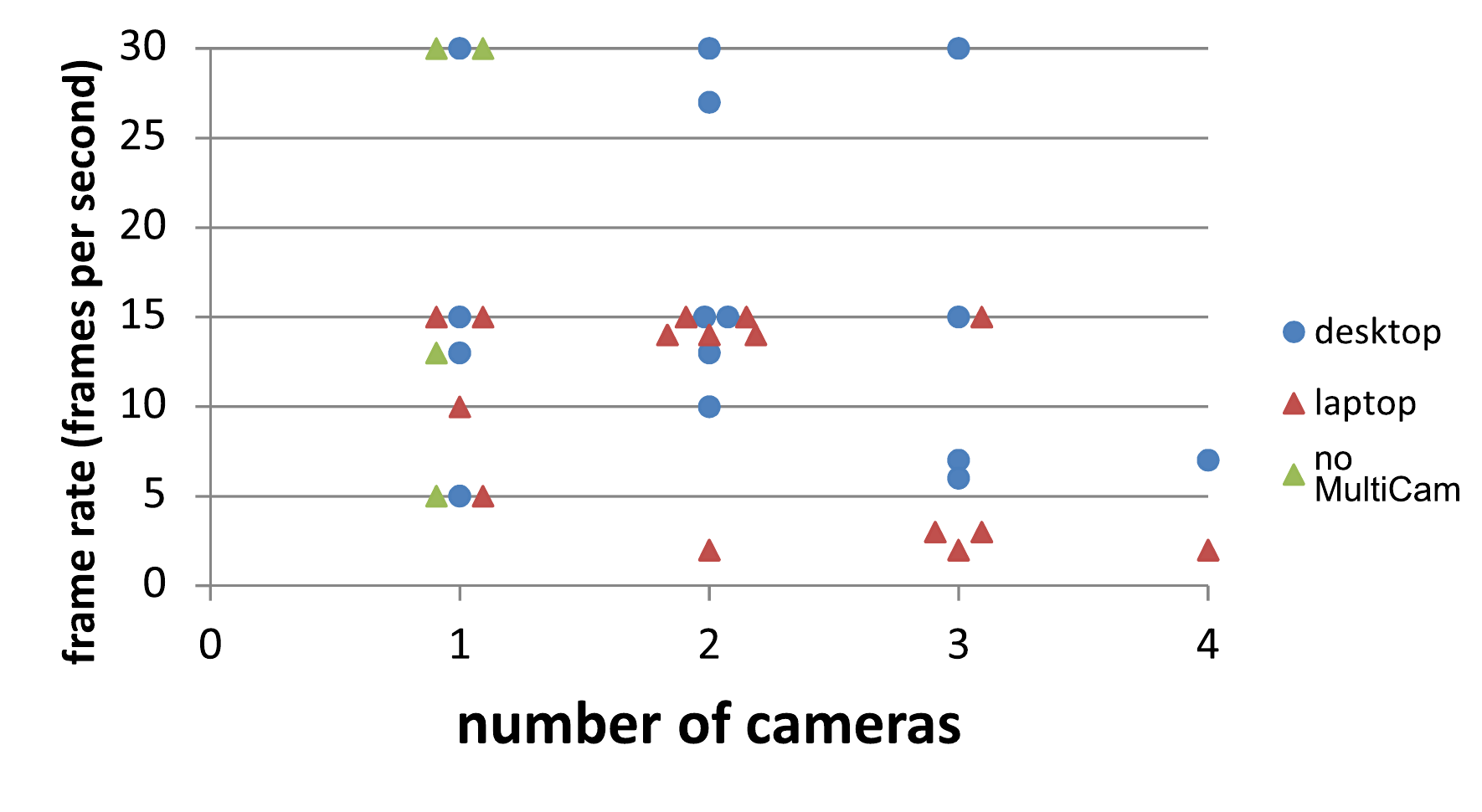} \\ [3mm]
    \includegraphics[width=\thiswidth]{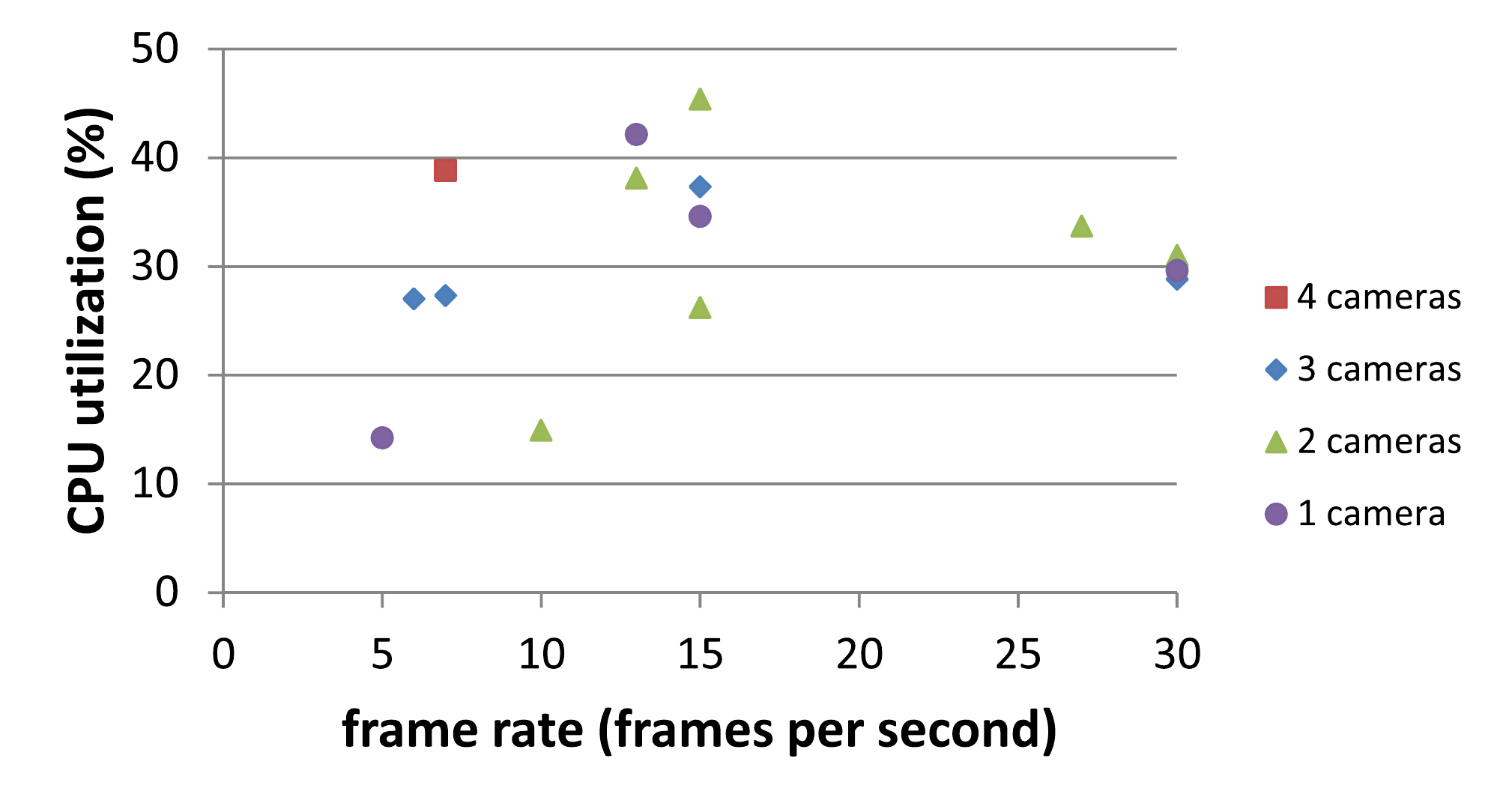} 
  \end{tabular}
  \caption{\textbf{MultiCam CPU utilization and frame rate during a
      Skype video chat.} Top: CPU utilization for all possible
    combinations of four different webcams, on two different
    computers.  Middle: Frame rate for the same set of experiment
    runs.  Bottom: The same data as the previous two graphs, combined
    a single graph (for clarity, only the runs on the desktop computer
    are shown).  In the top and middle panels, some points have been
    shifted horizontally to improve visibility.}
  \label{fig:cameras-CPU-skype}
\end{figure}

Figure~\ref{fig:cameras-CPU-skype} shows the same information as
Figure~\ref{fig:cameras-CPU}, but for the Skype runs rather than the
raw runs.  Comparing the top panels of Figures~\ref{fig:cameras-CPU}
and~\ref{fig:cameras-CPU-skype}, we see that Skype adds significant
CPU overhead to the local display benchmark. Presumably, this overhead
is primarily due to Skype's proprietary compression and encryption,
which have been analyzed in several prior works
(e.g. \cite{baset06skype,Zhang2012}).  Interestingly, the $3\times$
multiplier between desktop and laptop (which is expected based on CPU
specs, and held roughly true for the local display benchmark) is no
longer even approximately correct; the average multiplier is in fact
only 1.8.  But this discrepancy can be explained by the lower average
frame rate from the laptop: on average, the laptop frame rate was 1.6
times lower than the desktop, and since $1.6 \times 1.8 \approx 3.0$,
we recover the expected ratio of CPU utilization.

This brings us to the middle panel of
Figure~\ref{fig:cameras-CPU-skype}.  Comparing with
Figure~\ref{fig:cameras-CPU}, we see the desktop machine fared well
with the imposition of Skype.  In fact, 14~of the 15~camera
combinations had the same or similar frame rate
($\pm5$~fps).\footnote{The single exception here was the HD-3000,
  which achieves 29~fps on the local benchmark, but a puny 5~fps with
  Skype.}  But as already mentioned, the laptop suffered: six camera
combinations remained within $\pm5$~fps, but six others suffered
double-digit decreases.\footnote{And one combination improved by
  8~fps. Go figure!}  Also apparent from the middle panel of
Figure~\ref{fig:cameras-CPU-skype} is the lower frame rate achieved
for Skype runs by the laptop, compared with the desktop (this
contrasts with the raw runs).  Most likely, this is due to the lower
upload bandwidth of the laptop's location (recall that it employed
residential broadband).  A 2012 study~\cite{Zhang2012}, which used
Skype version~5.2, has shown that Skype monitors the available
bandwidth and congestion (probably via packet loss rate), and adjusts
its video codec as appropriate.  One of the codec's main adjustable
parameters is the frame rate, and it appears to target the discrete
values 5, 10, 15, and 30~fps---although as we see in the figure, this
may not always be achievable.  It remains an open question as to
whether Skype also moderates its codec to account for excessive CPU
usage.  Finally, this middle panel provides a sanity check, showing
data for using Skype with a single camera natively (i.e.\ without
MultiCam).  The same frame rate was achieved for three of the four
cameras, providing reasonable evidence that MultiCam itself is not
unduly hindering performance.


The bottom panel in Figure~\ref{fig:cameras-CPU-skype} shows the same
data as the previous two panels, but including only the desktop Skype
runs for clarity. As with the raw runs, we see a weak correlation
between frame rate and CPU usage.

The high-level message for the Skype runs
(Figure~\ref{fig:cameras-CPU-skype}) is the same as for the raw runs
(Figure~\ref{fig:cameras-CPU}): chatting with multiple cameras
simultaneously is feasible with a fraction of a typical machine's
resources, but performance can vary widely and sometimes mysteriously.
On the positive side, we again find (see the bottom panel) that one
combination of three cameras achieves 30~fps with modest CPU (30\% in
this case).  On the negative side, one camera used in isolation
languished at 5~fps on the desktop machine, and certain other
combinations of multiple cameras were almost as bad (see middle
panel).  Even worse, some of the laptop results were dreadful, with
frame rates as low as 2--3~fps.


\subsection{Experiment 2: Resource usage of other multi-camera software}
\label{sec:experiment2-other-software}

\begin{figure}
  \centering
    \renewcommand{\thiswidth}{140mm}
    \includegraphics[width=\thiswidth]{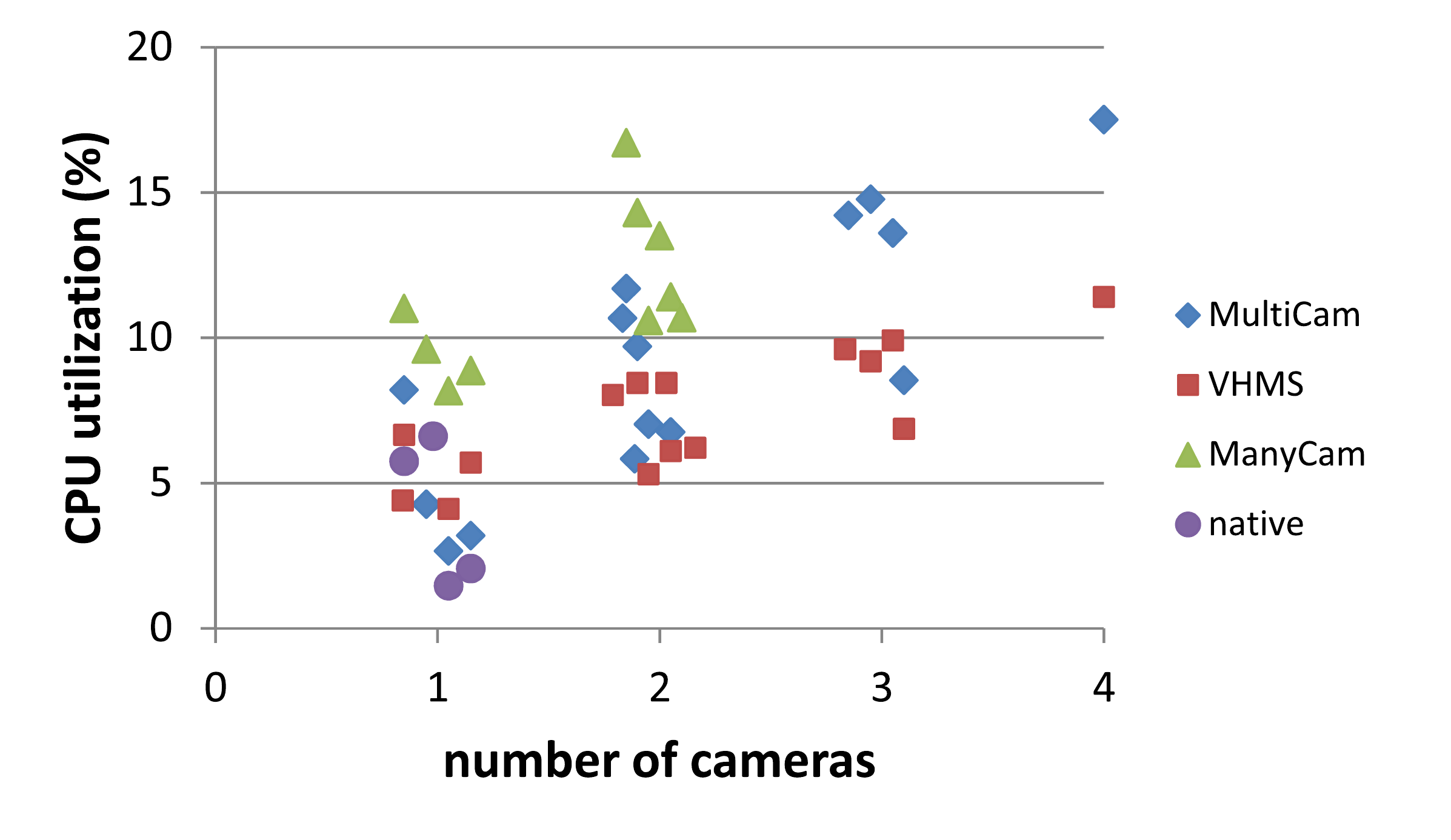}
    \caption{\textbf{CPU utilization for three multi-camera software
        systems.}  Points have been shifted horizontally to improve
      visibility.}
  \label{fig:CPU-competitors}
\end{figure}

It is natural to wonder if the resource consumption of MultiCam is
commensurate with other multi-camera software. Experiment~2 addresses
this by measuring the CPU utilization of MultiCam and two other
multi-camera systems: ManyCam, and VHMS (see
Section~\ref{sec:related-work} for descriptions of these systems).
Figure~\ref{fig:CPU-competitors} shows the results of running PlayCap
(the same local display benchmark as in Experiment~1) on the same
15~camera combinations, for each of the three systems. ManyCam
supports at most two cameras, so this system has no data points for
the 3- and 4-camera combinations. In addition, the benchmark was run
for each single camera without any virtual camera software; these are
the points labeled ``native'' in Figure~\ref{fig:CPU-competitors}.

The high-level conclusion from this experiment is that MultiCam is
reasonable in its CPU consumption.  It is a little more expensive
(perhaps 5\% of CPU for three or four cameras) than VHMS, a little
more expensive (1--3\% of CPU) for a single camera than the camera
natively, and a little less expensive than ManyCam (about 5\% of CPU).

\subsection{Experiment 3: Camera-switching latency of other multi-camera software}
\label{sec:experiment3-camera-switching-latency}

As we already saw in the results of the user study
(Section~\ref{sec:user-study}), low latency of the camera-switch
operation is important for positive user experiences.  Two of the
design decisions for MultiCam (simultaneous data retrieval from all
cameras, and fixed output resolution---discussed in
Sections~\ref{sec:switching-implementation} and~\ref{sec:resolution}
respectively) were made with the explicit goal of low-latency
camera-switching. Experiment~3 investigates whether MultiCam is
successful in reducing camera-switching latency below that of existing
multi-camera software.

The experiment deliberately eliminates network latency from
consideration.  Although it is an important component of the user's
experienced latency, network latency is the same for all systems, so
would only add unnecessary noise to a comparison of camera-switching
latency. Therefore, we measure camera-switching latency on PlayCap,
the same local display benchmark employed in the previous two
experiments. Specifically, we measure the time between the user's
request to switch cameras (issued via a mouse click in all cases
considered here) and the resulting switch of views in the display
window.  A screen capture tool\footnote{CaptureWizPro,
  \url{http://www.pixelmetrics.com/CapWizPro}} was used to record both
the mouse click and the PlayCap output window in a single
movie. Latencies were computed by manually single-stepping through
these screen-capture movies to find the frames in which the relevant
events occurred. The screen-capture movies had a temporal resolution
of about 22~ms, so the uncertainty in any single measurement is
$\pm11$~ms.  The latency figures are derived from the difference of
two such measurements, so the uncertainty in the latencies is
$2\times11$, or approximately $\pm22$~ms. The experiment employed the
same 2~cameras (the two LifeCams) in each test, with the user
switching from a primary view of the VX-3000 to the HD-3000.

\begin{figure}
  \centering
    \renewcommand{\thiswidth}{140mm}
    \includegraphics[width=\thiswidth]{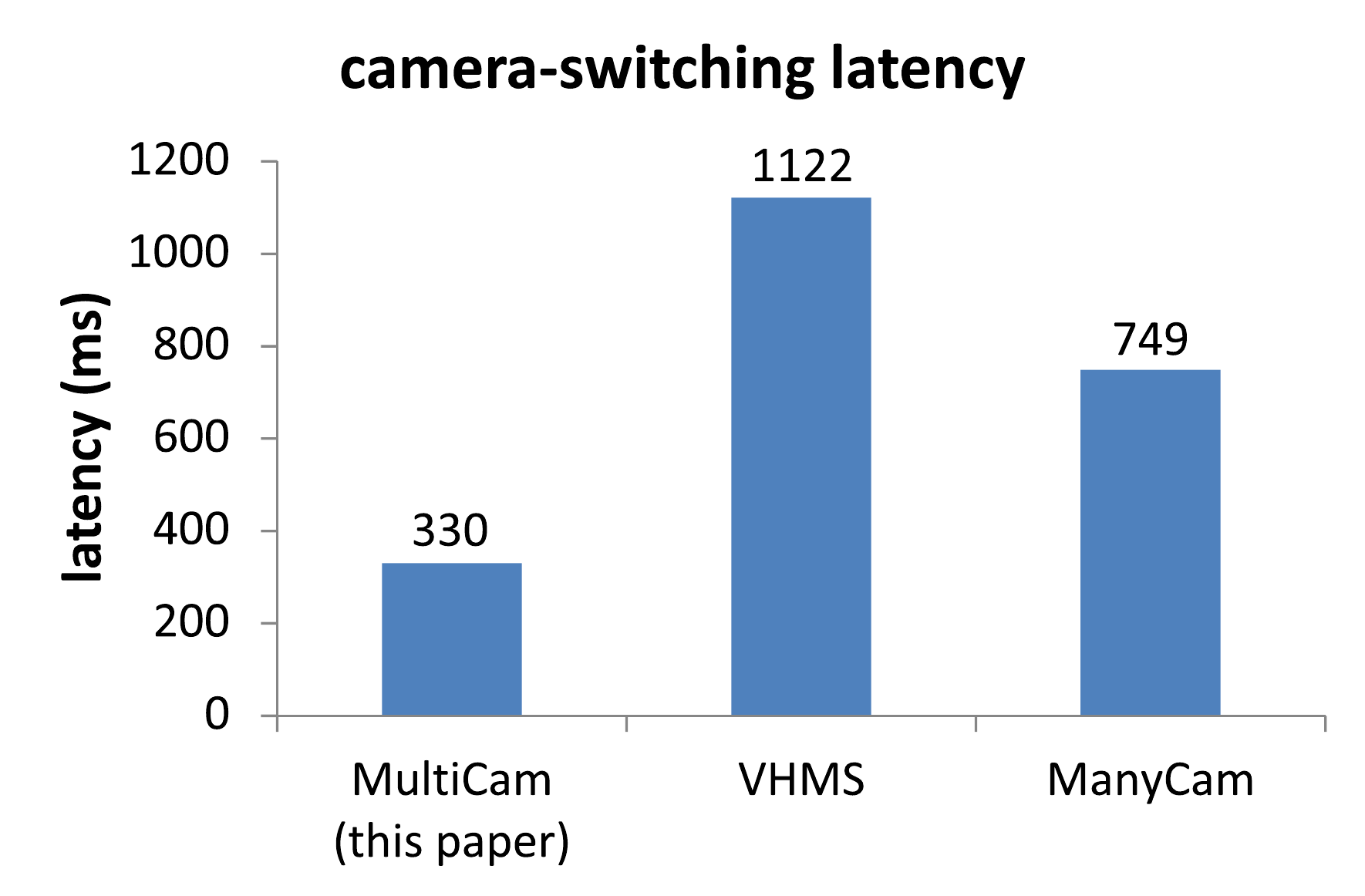}
    \caption{\textbf{Camera-switching latency for three multi-camera
        software systems.} }
  \label{fig:camera-switch-latency}
\end{figure}

Figure~\ref{fig:camera-switch-latency} shows the results for the same
systems compared in Experiment~2: this report's MultiCam, VHMS,
and ManyCam.  It is clear that MultiCam enjoys a significant advantage
here, being 2--3~times faster than the other systems. But note that
the true camera-switching latency experienced by a remote user will be
the sum of the amounts from Figure~\ref{fig:camera-switch-latency} and
the network latency, which may itself be hundreds of milliseconds.
Hence, the difference between the three systems observed by a remote
user would be less dramatic than this 2--3$\times$ factor.
Nevertheless, in order for the total switching latency to be tolerable
on a connection with non-negligible delay, it is clearly important for
the camera-switch itself to have low latency, and it seems MultiCam
has succeeded in achieving this.

The relatively large latency for the other two systems (more than one
second in the case of VHMS) suggests that these systems switch
cameras by stopping the running DirectShow graph, performing surgery
on it, then restarting the graph.  MultiCam's alternative approach of
keeping all cameras running and feeding only the relevant bits
downstream appears to save at least several hundred milliseconds, at a
modest cost in CPU, as we saw in Experiment~2.



\subsection{Experiment 4: Display latency of multiple cameras}
\label{sec:experiment4-multi-camera-switching}

It has been shown that for audio calls, Skype users' satisfaction is
much more strongly influenced by the transmitted signal's bitrate and
jitter than by its latency~\cite{chen06:skype}. But there do not appear
to be any similar results for video chat, so it seems desirable to
understand whether or not the simultaneous use of multiple cameras
affects video latency.  Experiment~4 investigates this.  As with
Experiment~3, network latency is eliminated---not because it is
unimportant, but because it is a constant added to any delay due to
multiple-camera use.  Therefore, we again consider latency for the
PlayCap local display benchmark.

\begin{figure}
  \centering
    \includegraphics{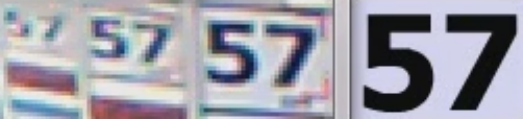}
    \caption{\textbf{Example of screen capture used to compute display
        latency.} The ``57'' on the right is displayed directly on the
      screen, and the reproductions of the ``57'' to the left are the
      result of three successive feedback loops through the camera.}
  \label{fig:latency-movie-still}
\end{figure}

The experimental method uses a type of recursion, inspired by
the method of vDelay~\cite{Boyaci09:skype}, but significantly
simplified since we are measuring local rather than remote display
latency.  The webcam whose latency is to be estimated is pointed at
the monitor where its own output is being displayed in the PlayCap
window.  Meanwhile, immediately adjacent to this window, a visible
event takes place at regular intervals---a counter incrementing once
per second was used in this case.  As in Experiment~3, a screen
capture tool is used, this time to capture the counter and any
relevant portion of the PlayCap window.
Figure~\ref{fig:latency-movie-still} shows the technique.  By
positioning the camera at the right angle, we can ensure that the
captured image of the counter is displayed immediately adjacent to the
counter itself (in this particular case, the original counter is on
the far right, and its image is immediately to the left). The captured
image of the image of the counter is also captured by the camera, and
displayed adjacent to the previous image. This process can be
continued indefinitely, but the image quality degrades with each trip
through the camera, and in practice the degradation was too severe to
permit analysis after the third iteration.    

By single-stepping through the resulting screen-capture movie, we can
determine the frame in which the rightmost and leftmost counters
increment.  The difference in the capture times of these frames is
equal to the camera's capture-to-display latency multiplied by the
number of iterations (which, as already mentioned, was three in most
cases).  The main advantage of using the feedback loop to measure
latency is accuracy. As with Experiment~3, the time resolution of the
screen capture is about 22~ms, but this uncertainty is reduced by a
factor of three when using the triple feedback loop.  For each camera
and scenario---described shortly---measurements were taken for about
10~separate events, and averaged to further reduce uncertainty.

Measurements were made for each of the four webcams used in the
previous experiments. More specifically, each camera's latency was
measured in two scenarios: (i) the given camera is the only one
connected to the MultiCam filter (the others might as well be
disconnected; they have no effect on the system), and (ii) all four
cameras are connected to the MultiCam filter and are simultaneously
displayed in tiled mode, but we measure the particular tile whose
content comes from the camera being measured.  The feedback system for
reducing uncertainty is not directly applicable to the tiled mode
scenario, so measurements of the tiled scenario were based off only a
single camera latency and have a correspondingly higher uncertainty.

\begin{figure}
  \centering
    \renewcommand{\thiswidth}{140mm}
    \includegraphics[width=\thiswidth]{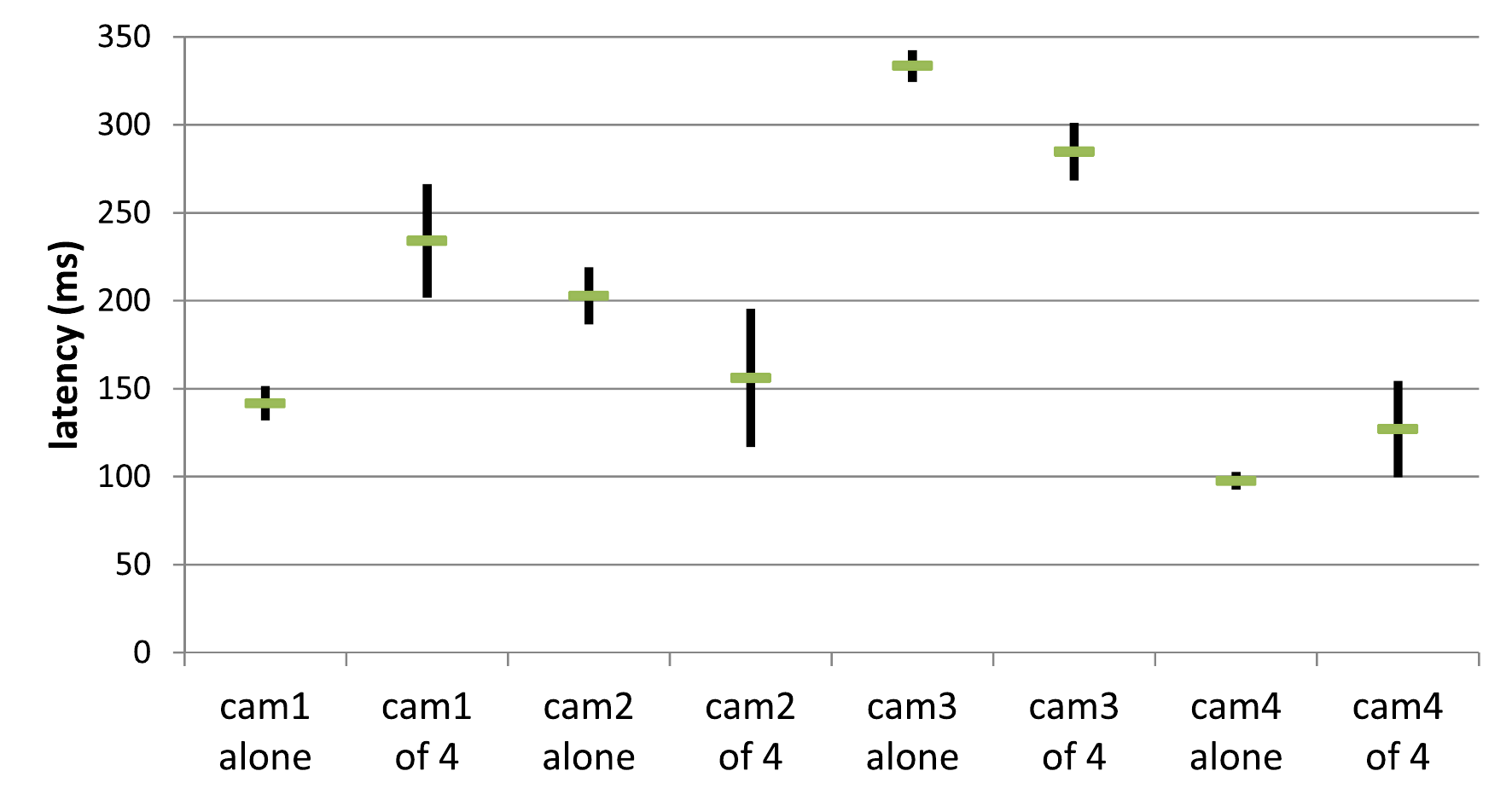}
    \caption{\textbf{MultiCam display latency for single and multiple
        cameras.} Horizontal lines show the mean and vertical lines
      show the standard deviation of the 10~latency measurements taken
      for each scenario. }
  \label{fig:display-latency}
\end{figure}

Figure~\ref{fig:display-latency} shows mean and standard deviation for
each camera and scenario.  For any given camera, we see a relatively
small difference between the single-camera and four-camera scenario;
two of these differences are increases and two are decreases.  Hence,
it seems safe to conclude that simultaneous use of up to four webcams
does not significantly alter the latency of video observed by video
chat users.  We can also see the dramatic differences in latencies
between different cameras---as much as a factor of 3, ranging from 100
to 300~ms.  This is yet another example of the ``chat user beware''
maxim emerging from these experiments. If video latency is an
important component of user satisfaction, then camera manufacturers
and video chat software developers should probably provide better
tools to help users choose appropriate technology.

\section{Privacy and invasiveness multi-camera video chat}
\label{sec:privacy-invasiveness}

The prospect of the listener controlling the view of the speaker's
location raises the question of whether the speaker's privacy
might be violated by such systems. As an extreme thought experiment,
suppose the speaker is in a room (such as an office or bedroom),
and employs a 360\textdegree\ panoramic, high-definition camera
stationed in the center of the room. Further suppose the listener can
zoom in on any portion of the camera's panorama, and the resolution is
sufficiently good that the listener could read the visible text of any
papers or books in view.  (Note that there is no consumer-grade
technology that achieves this at the time of writing; we are simply
using the scenario as an extreme case of potential privacy violation.
More realistically, we might imagine three or four cameras showing
different views of the room, with the listener switching between and
zooming in on just those available views.)  

Described in this way, the scenario feels somewhat creepy, and one is
tempted to immediately categorize it as a clear violation of
privacy. But this conclusion does not withstand more careful
scrutiny. Recall that our underlying objective is to restore to the
video chat participants the same freedoms they would have in a
face-to-face conversation.  Therefore, the privacy properties of the
system should be assessed by comparing with the privacy properties of
a face-to-face meeting in the same location as the multi-camera
system.  So, our thought experiment with a high-def panoramic camera
in the speaker's office or bedroom is only valid if that
speaker would have been comfortable having a face-to-face meeting
with the listener in the same office or bedroom. Obviously, if you
wouldn't invite your acquaintance Fred into your bedroom for a
face-to-face meeting, then you shouldn't chat with Fred via an
immersive video system based in your bedroom either.  On the other
hand, if you are comfortable with the idea of Fred wandering around
your bedroom or office, reading the titles of books on the wall and
glancing at the receipts on your desk, there is no rational reason to
avoid immersive video chat with Fred in that same location.

It might be argued that when Fred is physically present and looking at
your books or receipts, you get immediate feedback on his actions and
can take steps to prevent privacy violations (e.g.\ put the receipts
in a drawer as Fred walks over towards the desk).  In contrast, it
could be argued that a remote listener has more chance to snoop on
arbitrary parts of the environment without the speaker being
aware. There is some truth to this claim, especially given that the
video stream can be recorded and analyzed later, perhaps even using
super-resolution techniques~\cite{chaudhuri01:_super_resol_imagin}.
The obvious response is that the speaker should know the
capabilities of the camera and remove any private material from view
before the video chat starts. And the situation is further ameliorated
by the fact that the speaker can always monitor the listener's
view via the local view window (so if you see Fred is zooming in on
the receipts, put them away). Indeed, this is an important reason that
chat software should always provide a local view window option.

Overall, then, it is reasonable to conclude that immersive video chat
creates no privacy violations beyond the face-to-face conversations it
seeks to emulate.  But one of the fundamental points of this report is
that we can and should put an even more positive spin on the
situation. Provided the speaker has chosen a suitable location for the
video chat, it is actually a good thing that the listener can, for
example, examine books and pictures on the wall, see what kind of cell
phone the speaker has, and see the titles of some documents the
speaker happens to be reading.  These are exactly the kind of
serendipitous observations that lead to interesting and varied
conversations, and which are completely absent from today's video chat
paradigm.

\section{Discussion and future work}
\label{sec:discussion}

The most obvious opportunity for future work is to incorporate
non-standard imaging devices, such as panoramic cameras, into the mix
of cameras.  Improving the UI for camera-switching should be a
priority, perhaps using image stitching~\cite{Brown:2007:ijcv} to
combine views, or navigation between views inspired by the Photo
Tourism of Snavely \emph{et al.}~\cite{Snavely2006}.

This report has also highlighted some areas in which webcam
manufacturers and video chat software developers could improve the
multi-camera chat experience. These include
\begin{itemize}
\item addressing the unpredictable CPU usage, frame rate, and latency
  of cameras identified by the benchmark experiments
  (Section~\ref{sec:benchmarks})
\item standardizing a protocol for remote camera switching
  (Appendix~\ref{sec:MultiCam-Ap2Ap} provides specific suggestions)
\item providing cameras with a ``ready'' mode, whereby they can begin
  transmitting video data upon request, essentially instantaneously
\item providing features to distinguish between physical and virtual
  cameras (as discussed in Appendix~\ref{sec:distinguishing-virtual})
\item ensuring the current camera device is released before a
  newly-selected device is activated (see
  Appendix~\ref{sec:difficulty-switching})
\end{itemize}


\section{Conclusion}

Multi-camera video chat seems to be a promising and underutilized tool
in the multimedia milieu.  This report has demonstrated the feasibility
of multi-camera chat on standard consumer hardware, and suggested
scenarios in which multiple cameras improve the chat experience. A
user study provided strong empirical findings on the advantages and
disadvantages of listener-controlled switching between camera views.
Some design trade-offs inherent in multi-camera chat software were
discussed, and the report also presented MultiCam, an open-source
package providing multi-camera chat.  Perhaps researchers, software
developers, and hardware designers can build on these ideas to provide
rich, easily-controlled, multi-view video chat experiences in the
future.

\appendix

\section{MultiCam design details}
\label{sec:mult-design-deta}

This appendix fills in a variety of technical details relating to the
design of MultiCam.

\subsection{Creation and destruction of MultiCam filter}
\label{sec:filter-creation}

DirectShow filters are \cpp\ objects.  (More precisely, they are
Microsoft COM objects, but the distinction will not be important in
what follows.) An important consequence of this is that video chat
software creates a given DirectShow filter when it is needed for a
video chat, uses it for the duration of the chat, then destroys the
filter at the chat's conclusion.  So our design cannot assume that a
MultiCam filter exists on the system at any particular time.  This
fact is particularly important for communication between the
MultiCam application and the MultiCam filter.

\subsection{Temporary DirectShow graph}
\label{sec:temporary-graph}

Webcam capabilities are enumerated via the
\texttt{IAMStreamConfig::GetStreamCaps()} DirectShow method for filter
pins.  This method may be called on a pin whose filter is not
currently connected to a graph. In fact, Skype does exactly this on
startup: before creating any DirectShow graphs, all cameras on the
system are asked to enumerate their capabilities.  MultiCam faces a
chicken-and-egg problem here, since it does not know its capabilities
until it has been connected to its upstream filters (the physical
cameras).  

The reasons for this are a little technical, and rely on
details that were swept under the rug in the earlier discussion of
selecting webcam capabilities (Section~\ref{sec:resolution}).
DirectShow has a facility termed ``intelligent connect,'' whereby
filters mutually negotiate a suitable media type when they are
connected, based on various preferences and requirements defined
within the filters. MultiCam takes advantage of this.  Specifically,
MultiCam relies on intelligent connect to determine the media type of
the connection to the physical camera that happens to occur first in
the operating system's enumeration of cameras.  This media type is
used as the output media type, with some fields changed to take
account of the resolution selected by the method of
Section~\ref{sec:resolution}.  

The advantage of doing this is that the MultiCam filter's output media
type is identical (in almost all fields) to at least one of the
physical cameras. And if the physical cameras all happen to be
identical, all connections will use exactly the same media type and no
unnecessary conversions are performed. The disadvantage is that it is
impossible to enumerate the MultiCam filter's capabilities without
first connecting the filter to its upstream cameras.  To work around
this, the MultiCam code creates a temporary DirectShow graph if it is
needed at capability enumeration time, and destroys this graph as soon
as the enumeration is done.

\subsection{Distinguishing virtual cameras from physical cameras}
\label{sec:distinguishing-virtual}

MultiCam is targeted at consumers, and should ideally require no
configuration.  In particular, users should not be required to specify
the exact set of physical cameras to be used within
MultiCam---although advanced users should of course have the option to
specify the cameras if desired.  Hence, the default behavior, when a
MultiCam filter is instantiated, is for all physical cameras connected
to the system to be employed as inputs to the MultiCam filter.
Unfortunately, this ideal behavior does not appear to be achievable in
all cases. 

The problem is that there is no reliable way of distinguishing a
physical camera from a virtual camera.  Suppose, for example, that
ManyCam and MultiCam are both installed on a given system, which also
has two physical webcams connected.  When a MultiCam filter is
instantiated, the webcam drivers and the ManyCam driver appear to be
the same category of device.\footnote{Specifically, this is
  \VideoInputDeviceCategory.  See Appendix~\ref{sec:masq-as-phys} for
  details.}  So without using additional heuristics, MultiCam would
employ all three ``cameras'' as inputs.  This can have dire
results. For example, if the user had previously specified MultiCam as
an input to ManyCam, we would have a directed cycle in the DirectShow
graph, meaning it can't run successfully.  Undirected cycles can also
cause problems, as the following example shows.  Suppose the user had
previously specified one of the webcams---let's call it
Webcam~$A$---as the input to ManyCam.  When a MultiCam filter is
instantiated, both MultiCam and ManyCam attempt connections to
Webcam~$A$, but only one can succeed.  Note that an undirected cycle
does not always lead to an unrunnable graph.  In fact, one of the
chief features marketed for ManyCam is that it can be used
simultaneously as the input to multiple applications. But physical
cameras do not have this ability.

At present, MultiCam solves this problem as best it can using
heuristics.  All common virtual camera filters are automatically
excluded from the enumeration of camera devices.  Devices are also
excluded if they refuse to enumerate their capabilities or cannot be
connected to the graph in a timely fashion.  Finally, advanced users
can specify the exact set of devices to be used as inputs via a
configuration file. This is useful in several situations: (i) working
round situations where the heuristics fail; (ii) using a subset of the
physical cameras connected; (iii) deliberate use of a virtual camera
such as ManyCam as one of the inputs (which could be desirable for an
advanced user who has taken care not to create any cycles in the
DirectShow graph).

Although this problem is not severe, it's worth noting that programs
like MultiCam could be made more user-friendly if operating systems
could definitively distinguish between physical and virtual cameras.
This is a feature that could be included in future multimedia
frameworks.

\subsection{Difficulty switching between physical cameras and MultiCam}
\label{sec:difficulty-switching}

The present design of MultiCam has a subtle flaw that should be
acknowledged.  To explain this, suppose that a user is running Skype
and has previously selected a physical camera as the video input
device.  The user decides to switch to MultiCam as the input device,
and selects MultiCam in the Skype UI.  Unfortunately, it turns out
that Skype does not release the old device before activating the new
one.  This prevents MultiCam from starting properly, since (as
described above) MultiCam needs to activate each physical camera
itself before it can report its own capabilities. There are, no doubt,
relatively simple fixes for this problem, but they have not yet been
investigated. Indeed, ManyCam does not exhibit this symptom, which
proves that it is possible for a virtual camera to interact
satisfactorily with this part of the Skype UI.  On the other hand, the
existing design of MultiCam has no problems with Yahoo Messenger; in
that program, users can switch directly between MultiCam and a
physical camera. Hence, it seems worth mentioning this issue so that
Skype developers can address it if desired.

The current workaround for selecting MultiCam as the video input in
Skype is ugly but acceptable.  When MultiCam is installed, a separate
virtual camera, named \emph{VCam}, is also installed. VCam is a pure
virtual camera: it does not use any physical cameras as input, instead
generating random colors at every pixel in its output.  Hence, Skype
is perfectly happy for users to switch between a physical camera and
VCam, and between VCam and MultiCam.  So the workaround is for users
to temporarily select VCam as the input device whenever switching
between a single physical camera and MultiCam.

\section{Masquerading as a physical camera}
\label{sec:masq-as-phys}

To the best of my knowledge, there is no officially-published standard
for enumerating the devices suitable for video chat on a Windows
box. However, DirectShow filters may be registered as belonging to a
particular \emph{category} of filters (identified by a GUID), and a
widely-accepted de facto requirement is that video devices should be
registered in the \VideoInputDeviceCategory
filter category. For brevity, we will refer to this as the
\emph{VideoInputDevice} category.\footnote{There are at least two
  concrete pieces of evidence that membership of the VideoInputDevice
  category is indeed a de facto requirement for video devices. First,
  the GraphEdit utility provided by Microsoft as part of the
  DirectShow framework appears to list precisely these devices as ``Video
  Capture Sources.'' Second, Skype appears to ignore any device not in
  this category.}  Note that membership of the VideoInputDevice
category is a necessary, but not sufficient, condition for a device to
be usable in video chat. For example, the filter for a TV tuner is
likely to be registered in this category, but it may not be desirable
to offer the TV signal as an input to video chat.

Thus, merely registering a filter as a VideoInputDevice is not
sufficient. The filter must also behave sufficiently like a physical
camera, in the sense that it gives a sensible response to any
DirectShow API method call made by the video chat software.  Again,
there appears to be no published standard for this required
behavior. Presumably, every video chat application defines its own
standard implicitly, by running a battery of tests on each
VideoInputDevice to determine its suitability.  I am not aware of any
video chat application that documents these tests, and in the case of
proprietary software such as Skype, we cannot examine the source code
to check what the requirements might be.

Therefore, the following rather tedious, but effective, approach was
adopted in reverse engineering the behavior required of a virtual
camera suitable for Skype.  (The approach is described here in case it
is of use to other implementers of similar systems.)  First, we start
with the example code for the source filter \texttt{CSource} provided
by Microsoft.  As already discussed, ensure that the filter will be
registered as a VideoInputDevice.  Next, alter and add to this code by
implementing every virtual function in the class hierarchy above, and
including, \texttt{CSource}.  Each function should log the fact that
it was called and perhaps additional information about its parameters
and return values.  Now run the video chat software, choosing this new
filter as the video source. By examining the log, we can determine
which methods were called and make sensible guesses as to the desired
behavior.

There is a further complication which has been ignored up until this
point. In DirectShow, filters are connected via software abstractions
called \emph{pins}, which are implemented as \cpp\ classes derived from
a suitable base class.  A filter's pins determine much of its
behavior, so the technique described above for reverse engineering the
behavior of the virtual camera's filter must also be applied to the
virtual camera's output pin.  That is, derive a new class from
\texttt{CBaseOutputPin}, implement all virtual functions in the class
hierarchy, add suitable logging for every function, and test with
Skype.

\subsection{Skype's camera requirements}
\label{sec:skyp-camera-requ}

Here we briefly present the results of the reverse engineering
approach just described.  Of course, these results depend on
undocumented behavior of Skype, and this behavior could change with
future versions of Skype. Nevertheless, a quick summary of the results
(obtained for Skype version 5.5) may be a useful guide for others
implementing similar systems.  We assume the VideoInputDevice filter
is derived from \texttt{CTransformFilter}, and the filter's output pin is
derived from \texttt{CTransformOutputPin}.  Moreover, it is essential that the
output pin implement the \texttt{IKsPropertySet} and \texttt{IAMStreamConfig}
interfaces.

On startup and/or when the user is altering video settings, Skype
causes the following DirectShow methods to be called on the
VideoInputDevice filter:
\begin{itemize}
\item \texttt{CBaseFilter::GetPinCount()}
\item \texttt{CBaseFilter::GetPin()}
\item \texttt{CBaseFilter::JoinFilterGraph()}
\item \texttt{CTransformFilter::DecideBufferSize()}
\item \texttt{CTransformFilter::CheckInputType()}
\item \texttt{CTransformFilter::CheckTransform()}
\end{itemize}
Similarly, the Skype startup or video settings code causes the
following calls of DirectShow methods on the VideoInputDevice filter's
output pin:
\begin{itemize}
\item \texttt{IKsPropertySet::Get()} (when the property
  \texttt{AMPROPERTY\_\-PIN\_\-CATEGORY} is requested, this should return
  \texttt{PIN\_\-CATEGORY\_\-CAPTURE})
\item \texttt{IAMStreamConfig::GetFormat()}
\item \texttt{IAMStreamConfig::GetNumberOfCapabilities()}
\item \texttt{IAMStreamConfig::GetStreamCaps()}
\item \texttt{IAMStreamConfig::SetFormat()}
\item \texttt{CBasePin::GetMediaType()}
\item \texttt{CBasePin::Connect()}
\item \texttt{CBasePin::CheckMediaType()}
\end{itemize}
Hence, all of the above methods must be implemented and return
sensible results in order for a virtual camera to masquerade as a
physical camera for use in Skype.

\section{MultiCam Ap2Ap protocol}
\label{sec:MultiCam-Ap2Ap}

As discussed in Section~\ref{sec:design-overview}, two instances of
the MultiCam application communicate using Skype's Ap2Ap facility,
which is part of the Skype desktop API.  The Ap2Ap facility allows
MultiCam instances to exchange arbitrary UTF-8-encoded strings.  Here,
we describe the strings actually used. This collection of commands is
termed the \emph{MultiCam Ap2Ap protocol}.  The protocol consists of
four types of requests, three of which require responses.  Hence,
there are five commands in the protocol: Ping, Pong, AskNumCams,
ReplyNumCams, AskVersion, ReplyVersion and AdvanceCamera. These are
described in detail below, but let us first establish some
notation. If the protocol specifies that the string ``\texttt{FOO}''
should be sent, this means that ``\texttt{FOO}'' should be first
appended to an appropriate command in the Skype Desktop API.  A
typical example of a string actually sent to Skype would be:
\begin{quotation}
  \texttt{ALTER APPLICATION multicam WRITE skypeusername:1 FOO}
\end{quotation}
This example assumes that an Ap2Ap connection named \texttt{multicam}
has already been established with a communication stream named
\texttt{skypeusername:1}.  Please see the documentation of the Skype
Desktop API~\cite{SkypeAPI2011} for details of how to achieve this.  

The existing implementation of MultiCam assumes an Ap2Ap connection
named \texttt{multicam} is used (as in the example above), but it is
possible to envisage the same MultiCam Ap2Ap protocol being used by
other third-party applications using different Ap2Ap connection names,
so this requirement is not part of the protocol specification.

\subsection{Ping request and Pong response}
\label{sec:ping-request}

A Ping request consists of the string \texttt{AP2AP\_PING}.  On
receiving this request, MultiCam should respond with the string
\texttt{AP2AP\_PONG}.  These commands are obviously useful for
debugging, but they are also used in the implementation to check for
the presence of a remote MultiCam instance.

\subsection{AskNumCams request and ReplyNumCams response}
\label{sec:askn-requ-replyn}

An AskNumCams request consists of the string
\texttt{AP2AP\_ASK\_NUMCAMS}.  On receiving this request, MultiCam
should respond with a ReplyNumCams response.  The response consists of
the string of the form
\begin{quote}
  \texttt{AP2AP\_REPLY\_NUMCAMS\textvisiblespace n}
\end{quote}
where
`\textvisiblespace' represents a space character, and $n$ is the
number of physical cameras currently detected by the MultiCam filter
(or 0 if the filter is unavailable).  The number $n$ should be
formatted as the standard UTF-8 text representation of an integer.

The ReplyNumCams response may also be sent at any other time (i.e.\
without necessarily waiting for an AskNumCams request).  For example,
the current implementation sends ReplyNumCams when the MultiCam Ap2Ap
connection is first achieved.

\subsection{AskVersion request and ReplyVersion response}
\label{sec:askv-requ-replyv}

An AskVersion request consists of the string
\texttt{AP2AP\_ASK\_VERSION}.  On receiving this request, MultiCam
should respond with a ReplyVersion response.  The response consists of
a string of the form
\begin{quote}
  \texttt{AP2AP\_REPLY\_VERSION\textvisiblespace d.d\textvisiblespace
    d.d.d.d}
\end{quote}
where `\textvisiblespace' represents a space character, and each $d$
is a digit.  The first set of digits (``\texttt{d.d}'') is the version
number of the MultiCam Ap2Ap protocol (1.1, at the time of writing).
The second set of digits (``\texttt{d.d.d.d}'') is the version number
of the MultiCam application program (0.1.0.8, at the time of writing).

\subsection{AdvanceCamera request}
\label{sec:advanc-camera-requ}

An AdvanceCamera request consists of the string
\texttt{AP2AP\_ADVANCE\_CAMERA}.  There is no response.  On receiving
this request, the MultiCam application should send an AdvanceCamera
message to the MultiCam filter on the local machine (or, if the filter
is not available, do nothing).  As described in
Appendix~\ref{sec:AdvanceCamera-Ap2Filt}, this message will have the
effect of either switching the MultiCam filter between tiled and
non-tiled mode, or advancing the primary camera within non-tiled mode.

\section{MultiCam Ap2Filt protocol}
\label{sec:multicam-ap2filt-prot}

The MultiCam application and MultiCam filter communicate with each
other via the \emph{MultiCam Ap2Filt protocol}, described in this
section.  Ap2Filt messages are transmitted using the standard
Microsoft Windows messaging functionality, via Win32 API functions
such as \texttt{SendMessage()} and \texttt{SendMessageTimeout()}.  The
MultiCam application and MultiCam filter each create a hidden window
whose sole purpose is to send and receive these messages.  There are
seven message types: Discover, Attach, Kick, Ping, Pong,
AdvanceCamera, and Reset.  Each employs a different message ID (where
\emph{message ID} is defined as the second parameter in the Win32
\texttt{SendMessage()} function).  The message IDs themselves are not
fixed in advance, but are determined each time the protocol is
instantiated, using the Win32 \texttt{RegisterWindowMessage()}
function.  The parameters needed by \texttt{RegisterWindowMessage()}
are
\begin{quotation}
  \noindent
  \texttt{MulticamDiscover4AD2E57A-AF70-42AE-9A64-BC88F995B9C8}\\
  \texttt{MulticamAttach4AD2E57A-AF70-42AE-9A64-BC88F995B9C8}\\
  \texttt{MulticamAdvance4AD2E57A-AF70-42AE-9A64-BC88F995B9C8}\\
  \texttt{MulticamKick4AD2E57A-AF70-42AE-9A64-BC88F995B9C8}\\
  \texttt{MulticamPing4AD2E57A-AF70-42AE-9A64-BC88F995B9C8}\\
  \texttt{MulticamPong4AD2E57A-AF70-42AE-9A64-BC88F995B9C8}\\
  \texttt{MulticamReset4AD2E57A-AF70-42AE-9A64-BC88F995B9C8}
\end{quotation}
respectively, for each of the seven message types.  

The family of Win32 messaging functions such as \texttt{SendMessage()}
all employ similar parameters, including the \texttt{hWnd},
\texttt{wParam}, and \texttt{lParam} parameters which are referenced
in the discussion below.  For more information about the data types
and usage of these parameters, see the Win32 API
documentation~\cite{WindowsAPI}.

\subsection{Discover, Attach, and Kick Ap2Filt messages}
\label{sec:disc-attach-kick}

The Discover message is broadcast by the MultiCam filter whenever an
instance of the filter is created. Here, ``broadcast'' means that the
\texttt{hWnd} parameter is \texttt{HWND\_BROADCAST}.  The
\texttt{wParam} parameter is set to the handle of the filter's hidden
window, and the \texttt{lParam} parameter is set to the number of
physical cameras connected to the filter.

If there is no MultiCam application present when the Discover message
is broadcast, the message has no effect. If a MultiCam application is
present, it should attach itself to the filter. Specifically, the
MultiCam application sends an Attach message directly to the filter's
hidden window (it can do this as it has just received that window's
handle), setting the \texttt{wParam} parameter to the handle of its
own hidden window.  On receiving the Attach message, the MultiCam
filter stores the application's hidden window's handle for later use.

The Kick message is used by the MultiCam application in order to
kickstart the Discover-Attach sequence just described. This is needed
because sometimes a user will start the MultiCam application only
after the MultiCam filter has already been created by the video chat
software.  A Kick can also be used to restart communication after an
unexpected breakdown.  The usage of Kick is entirely straightforward:
the MultiCam application broadcasts the message (i.e.\ \texttt{hWnd}
is set to \texttt{HWND\_BROADCAST}), setting the \texttt{wParam} and
\texttt{lParam} parameters to arbitrary values since they will be
ignored.  On receiving a Kick, the MultiCam filter sends a Discover
message as described above.

\subsection{Ping and Pong Ap2Filt messages}
\label{sec:ping-pong-Ap2Filt}

Ping and Pong messages are used for debugging and for periodically
checking the connection between the MultiCam application and filter.
Specifically, the MultiCam application sends a Ping to the MultiCam
filter whenever it chooses (the \texttt{wParam} and \texttt{lParam}
parameter values are irrelevant).  On receiving a Ping, the MultiCam
filter sends a Pong message back to the MultiCam application (again,
the \texttt{wParam} and \texttt{lParam} parameter values are
irrelevant).

\subsection{AdvanceCamera Ap2Filt message}
\label{sec:AdvanceCamera-Ap2Filt}

The AdvanceCamera message is sent from the MultiCam application to the
MultiCam filter. Its purpose is to either switch to a new primary
camera (in non-tiled mode) or to switch between tiled and non-tiled
modes.  The \texttt{wParam} and \texttt{lParam} parameter values are
not used in this message.

Note that the physical cameras connected to the MultiCam filter have a
particular ordering which is established when the filter first uses
the DirectShow API to enumerate the cameras. This enables us to define
the behavior of the MultiCam filter when it receives an AdvanceCamera
message as follows. If the filter is currently in non-tiled mode, and
the primary camera is not the last camera, the filter remains in
non-tiled mode and new primary camera is the successor of the old
primary camera.  If the filter is in non-tiled mode, and the primary
camera \emph{is} the last camera, the filter switches to tiled
mode. If the filter is in tiled mode, it switches to non-tiled mode
and sets the primary camera to be the first camera.

\subsection{Reset Ap2Filt message}
\label{sec:reset-Ap2Filt}

The Reset message is sent from the MultiCam application to the
MultiCam filter.  Its purpose is to force a reinitialization of the
MultiCam filter. This may be useful if an unexpected error is
encountered, or if an unhandled change in the hardware configuration
(such as the addition or removal of a physical camera) has occurred.
The \texttt{wParam} and \texttt{lParam} parameter values are not used
in this message.  The precise behavior of the MultiCam filter on
receiving a Reset message is not specified here.  Note that the
current implementation does not implement this feature, but it is
anticipated that future versions of MultiCam may benefit from it.

\section{User study questionnaire}
\label{sec:user-study-questionnaire}

The following survey was administered verbally to each participant in
the user study described in Section~\ref{sec:user-study}.

\begin{enumerate}
\item Overall, which do you feel gave you a more satisfactory
  experience: you, the listener, controlling the cameras; or me, the
  speaker, controlling the cameras?  If you have no preference, that
  is also a valid answer.
\item $[$Skip this question if no preference is expressed on the
  previous question.$]$ This question asks how strongly you feel about
  your answer to the previous question. I will read a statement
  summarizing your answer to the previous question, and then ask you
  to tell me whether you (i) strongly agree, (ii) agree, or (iii)
  mildly agree with the statement. ``When the listener/speaker
  controlled the camera, the overall experience was more
  satisfactory.'' $[$Choose ``listener'' or ``speaker'' in this
  statement according to the answer to the previous question.$]$
  Please tell me whether you strongly agree, agree, or mildly agree
  with that statement.
\item When I, the speaker, was controlling the camera, was there anything
  that you liked or disliked about that experience?  List as many
  things as you wish.
\item When you, the listener, were controlling the camera, was there
  anything that you liked or disliked about that experience?  List as
  many things as you wish.
\item When you were controlling the camera, did you use the
  simultaneous view of both cameras much, if at all?
\item Please tell me any other thoughts or feelings you have about
  your experience using this system.
\end{enumerate}

\bibliographystyle{plain}

\end{document}